\documentclass{emulateapj}
\usepackage{natbib}
\usepackage{amsmath}

\begin{document}
\title{A Universal Mass Profile for Dwarf Spheroidal Galaxies?\footnote{This paper presents data gathered with the Magellan Telescopes at Las Campanas Observatory, Chile, and the MMT Observatory at Mt. Hopkins, Arizona.}}
\shorttitle{Universal Mass Profile for dSphs}
\author{Matthew G. Walker\altaffilmark{1}, Mario Mateo\altaffilmark{2}, Edward W. Olszewski\altaffilmark{3}, Jorge Pe\~narrubia\altaffilmark{1},  \\
N. Wyn Evans\altaffilmark{1} and Gerard Gilmore\altaffilmark{1}}
\email{walker@ast.cam.ac.uk}
\altaffiltext{1}{Institute of Astronomy, University of Cambridge, UK}
\altaffiltext{2}{Department of Astronomy, University of Michigan, Ann Arbor}
\altaffiltext{3}{Steward Observatory, The University of Arizona, Tucson, AZ}

\begin{abstract} 

We apply the Jeans equation to estimate masses for eight of the brightest dwarf spheroidal (dSph) galaxies.  For Fornax, the dSph with the largest kinematic data set, we obtain a model-independent constraint on the maximum-circular velocity, $V_{max}=18_{-3}^{+5}$ km s$^{-1}$.  Although we obtain only lower-limits of $V_{max}\ga 10$ km s$^{-1}$ for the remaining dSphs, we find that in all cases the enclosed mass at the projected half-light radius is well constrained and robust to a wide range of halo models and velocity anisotropies.  We derive a simple analytic formula that estimates $M(r_{half})$ accurately with respect to results from the full Jeans analysis.  Applying this formula to the entire population of Local Group dSphs with published kinematic data, we demonstrate a correlation such that $M(r_{half})\propto r_{half}^{1.4\pm 0.4}$, or in terms of the mean density interior to the half-light radius, $\langle \rho\rangle \propto r_{half}^{-1.6\pm 0.4}$.  This relation is driven by the fact that the dSph data exhibit a correlation between global velocity dispersion and half-light radius.  We argue that tidal forces are unlikely to have introduced this relation, but tides may have increased the scatter and/or altered the slope.  While the data are well described by mass profiles ranging over a factor of $\la 2$ in normalization ($V_{max}\sim 10-20$ km s$^{-1}$), we consider the hypothesis that all dSphs are embedded within a ``universal'' dark matter halo.  We show that in addition to the power law $M\propto r^{1.4}$, viable candidates include a cuspy ``NFW'' halo with $V_{max}\sim 15$ km s$^{-1}$ and scale radius $r_0\sim 800$ pc, as well as a cored halo with $V_{max}\sim 13$ km s$^{-1}$ and $r_0\sim 150$ pc.  Finally, assuming that their measured velocity dispersions accurately reflect their masses, the smallest dSphs now allow us to resolve dSph densities at radii as small as a few tens of pc.  At these small scales we find mean densities as large as $\langle \rho\rangle \la 5 M_{\odot}$pc$^{-3}$ ($\la 200$GeV cm$^{-3}$). 
\end{abstract}


\section{Introduction}

Dwarf spheroidal (dSph) galaxies are the smallest and least luminous galaxies in the Universe, and so provide unique diagnostics of galaxy formation at small scales.  Early observations (e.g., \citealt{aaronson83}) of stellar kinematics in the brightest ($L_V\sim 10^{5-7}L_{V,\odot}$) of the Milky Way's dSph satellites yielded the surprising result that all well-studied dSphs have central velocity dispersions of $\sim 10$ km s$^{-1}$, larger than expected for self-gravitating, equilibrium stellar populations with scale radii of $\sim 100$ pc.  Following work showing that such dispersions are unlikely to have been inflated significantly by unidentified binaries \citep{edo96,hargreaves96b}, stellar-atmospheric turbulence (e.g., \citealt{pryor88}) or tidal disruption \citep{pp95,oh95}, it has become widely accepted that dSphs are dominated by dark matter, with mass-to-light ratios of $M/L_V\sim 10^{1-2}$ (\citealt{mateo98} and references therein).  Furthermore, that dSphs have similar velocity dispersions despite varying over several orders of magnitude in luminosity suggests that the dark matter halos of dSphs are more similar than the differences in their stellar components might otherwise suggest.  Along these lines, \citet{mateo93} noted that all dSphs measured at that time had dynamical masses of $\sim 10^7M_{\odot}$, implying an anti-correlation between luminosity and $M/L$ (see, e.g., Figure 9 of \citealt{mateo98}).  
The recent discoveries of ultra-faint Milky Way satellites (e.g., \citealt{willman05b,zucker06a,zucker06b,belokurov07}) extend the range of dSph structural parameters by an order of magnitude in (luminous) scale radius and by three orders of magnitude in luminosity.  These objects provide new opportunities to identify scaling relations that may point to unifying principles that govern dSph formation and evolution.  \citet[``S08'' hereafter]{strigari08} present  one such (non-) relation, sharpening the point made by \citet{mateo93} by arguing that, despite spanning five orders of magnitude in luminosity, all the studied dSphs have mass $\sim 10^7M_{\odot}$ enclosed within their central $300$ pc.  S08 contend that a common value of $M_{300}\sim 10^7M_{\odot}$ may represent the minimum mass of a dark matter halo, or it may represent the minimum mass required for galaxy formation (e.g., \citealt{li09,maccio09}).

Here we present an analysis that differs from that of S08 in one key respect.  Rather than $M_{300}$, we consider the mass enclosed at the half-light radius, $M(r_{half})$, which is well constrained by the available kinematic data.  Our reasons for this choice all stem from the fact that, among the Local Group's known dSphs, $r_{half}$ varies over two orders of magnitude, from a few tens of pc to $\sim 1$ kpc \citep{ih95,mcconnachie06,martin08}.  First, regarding the faintest dSphs for which all detected members lie within projected radii $R \leq 100$ pc, $M_{300}$ has literal meaning only if one extrapolates the properties of the inferred dark matter halo far beyond regions sampled by observational data.  Since there is no evidence that the dark matter halos of the smallest systems extend to radii of order 300 pc, for these objects $M_{300}$ is merely a parameter setting the normalization of the mass profile (Section \ref{subsec:normalization}).  Second, a common value of $M_{300}$ among dSphs relays little information about their total masses---e.g., one can build a plausible Milky Way model using a central halo with $V_{max}\sim 200$ km s$^{-1}$ and $M_{300}\sim 10^7M_{\odot}$.  Third, for the smallest galaxies, constraints on $M(r_{half})$ imply constraints on dark matter densities in regions as small as $r\sim 30$ pc, providing unprecedented mass resolution on small galactic scales.  Finally, by considering the empirical relation we derive between $M(r_{half})$ and $r_{half}$, we find reason to make a stronger claim than S08's concerning the common mass of dSph satellites.  Specifically, we find that the entire population of Local Group dSphs can be fit reasonably well with a ``universal'' dark matter halo, about which the scatter with respect to the data is similar to that about a common value of $M_{300}$.  That is, to the extent that dSphs can be said to have a common $M_{300}$, the data suggest dSphs may have common masses at all radii.  

\section{Data}

The analysis presented below proceeds in two stages.  In the first (Section \ref{sec:jeans}), we apply the Jeans equation to estimate halo parameters of the eight dSphs for which we have acquired large kinematic data sets.  For the Carina, Fornax, Sculptor and Sextans dSphs we use kinematic data obtained with the Michigan/MIKE Fiber Spectrograph (MMFS) at Magellan.  \citet{walker07a} describe the acquisition and reduction of these data, while \citet{walker09a} present the entire data set.  For Draco, Leo I, Leo II and Ursa Minor we use data gathered with the Hectochelle fiber spectrograph at the MMT.  \citet{mateo08} describe the acquisition and reduction of these data and present the data for Leo I.  Papers presenting Hectochelle data for the remaining three galaxies are in preparation.  

One purpose of our Jeans analysis is to justify the use of a simpler method of estimating enclosed masses (Section \ref{subsec:simple}), which we can then apply to a broader sample of objects.  In the second stage of this work (Section \ref{sec:scaling}), we demonstrate correlations among the bulk structural and kinematic properties of dSphs.  There we benefit by considering the entire dSph population, including not only the bright Milky Way satellites but also the ultra-faint satellites discovered (e.g., \citealt{willman05a,zucker06a,zucker06b,belokurov07}) and observed spectroscopically (e.g., \citealt{martin07,simon07,belokurov09}) during the past five years.  We also include three dSph satellites of M31 (AndII, AndIX and AndXV) that have published kinematic data (we do not include AndXVI, for which \citep{letarte09} report an unresolved velocity dispersion), as well as the two remote dSphs (Tucana and Cetus) at the Local Group's outskirts.  For all of these objects, Table \ref{tab:dsphs} lists (projected) half-light radii, velocity dispersions and luminosities that we adopt from either the literature or our own data.  

In the interest of uniformity, we have chosen where possible to adopt measurements from studies of multiple objects observed and analyzed using the same methodology.  We caution that in some cases this preference may introduce systematic bias.  For example, for most of the low-luminosity dSphs we adopt the structural parameters measured from SDSS data by \citet{martin08}.  In our experience, deeper follow-up imaging for the faintest objects often indicates smaller half-light radii than are apparent from SDSS data alone---compare, for example, the stellar maps of Segue 2 from SDSS and MMT/Megacam imaging in Figure 3 of \citet{belokurov09} (see also the LBT follow-up imaging of Hercules by \citealt{sand09}).  In most cases the difference is slight, but we caution that the structural parameters of the faintest systems may be subject to revision when deeper data become available for more of these objects.

\begin{deluxetable*}{lrrrrrr}
\tabletypesize{\scriptsize}
\tablewidth{0pc}
\tablecaption{dSph Structural Parameters, Velocity Dispersions and Estimated Masses\tablenotemark{*}}
\tablehead{\\
  \colhead{Object}&\colhead{$L_V$}&\colhead{$r_{\mathrm{half}}$}&\colhead{$\sigma_{V_0}$}&\colhead{$M(r_{\mathrm{half}})$}&\colhead{$\langle \rho\rangle$}&\colhead{Ref.\tablenotemark{**}}\\
  \colhead{}&\colhead{$[L_{V,\odot}]$}&\colhead{[pc]}&\colhead{[km s$^{-1}$]}&\colhead{$M_{\odot}$}&\colhead{$M_{\odot}$pc$^{-3}$}\\
}
\startdata
\\
Carina&$2.4\pm 1.0\times 10^5$&$241\pm 23$&$6.6\pm 1.2$&$6.1\pm 2.3\times 10^6$&$1.0\pm 0.4\times 10^{-1}$&1,2\\
Draco&$2.7\pm 0.4\times 10^5$&$196\pm 12$&$9.1\pm 1.2$&$9.4\pm 2.5\times 10^6$&$3.0\pm 0.8\times 10^{-1}$&3,4\\
Fornax&$1.4\pm 0.4\times 10^7$&$668\pm 34$&$11.7\pm 0.9$&$5.3\pm 0.9\times 10^7$&$4.2\pm 0.7\times 10^{-2}$&1,2\\
Leo I&$3.4\pm 1.1\times 10^6$&$246\pm 19$&$9.2\pm 1.4$&$1.2\pm 0.4\times 10^7$&$1.9\pm 0.6\times 10^{-1}$&1,5\\
Leo II&$5.9\pm 1.8\times 10^5$&$151\pm 17$&$6.6\pm 0.7$&$3.8\pm 0.9\times 10^6$&$2.6\pm 0.6\times 10^{-1}$&1,6\\
Sculptor&$1.4\pm 0.6\times 10^6$&$260\pm 39$&$9.2\pm 1.1$&$1.3\pm 0.4\times 10^7$&$1.7\pm 0.5\times 10^{-1}$&1,2\\
Sextans&$4.1\pm 1.9\times 10^5$&$682\pm 117$&$7.9\pm 1.3$&$2.5\pm 0.9\times 10^7$&$1.9\pm 0.7\times 10^{-2}$&1,2\\
UMi&$2.0\pm 0.9\times 10^5$&$280\pm 15$&$9.5\pm 1.2$&$1.5\pm 0.4\times 10^7$&$1.6\pm 0.4\times 10^{-1}$&1,7\\
\\
Bootes 1&$3.0\pm 0.6\times 10^4$&$242\pm 21$&$6.5\pm 2.0$&$5.9\pm 3.7\times 10^6$&$1.0\pm 0.6\times 10^{-1}$&3,8\\
Bootes 2&$1.0\pm 0.8\times 10^3$&$51\pm 17$&$10.5\pm 7.4$&$3.3\pm 3.3\times 10^6$&$5.9\pm 5.9$&3,9\\
CVen I&$2.3\pm 0.3\times 10^5$&$564\pm 36$&$7.6\pm 0.4$&$1.9\pm 0.2\times 10^7$&$2.5\pm 0.3\times 10^{-2}$&3,10\\
CVen II&$7.9\pm 3.6\times 10^3$&$74\pm 12$&$4.6\pm 1.0$&$9.1\pm 4.2\times 10^5$&$5.3\pm 2.5\times 10^{-1}$&3,10\\
Coma&$3.7\pm 1.7\times 10^3$&$77\pm 10$&$4.6\pm 0.8$&$9.4\pm 3.5\times 10^5$&$4.9\pm 1.8\times 10^{-1}$&3,10\\
Hercules&$3.6\pm 1.1\times 10^4$&$330\pm 63$&$3.7\pm 0.9$&$2.6\pm 1.4\times 10^6$&$1.7\pm 0.9\times 10^{-2}$&3,11\\
Leo IV&$8.7\pm 4.6\times 10^3$&$116\pm 30$&$3.3\pm 1.7$&$7.3\pm 7.3\times 10^5$&$1.1\pm 1.1\times 10^{-1}$&3,10\\
Leo V&$4.5\pm 2.6\times10^3$&$42\pm 5$&$2.4\pm 1.9$&$1.4\pm 1.4\times 10^5$&$4.5\pm 4.5\times 10^{-1}$&12,13\\
Leo T&$5.9\pm 1.8\times 10^4$&$178\pm 39$&$7.5\pm 1.6$&$5.8\pm 2.8\times 10^6$&$2.5\pm 1.2\times 10^{-1}$&3,10,14\\
Segue 1&$3.3\pm 2.1\times 10^2$&$29\pm 7$&$4.3\pm 1.2$&$3.1\pm 1.9\times 10^5$&$3.0\pm 1.8$&3,15\\
Segue 2&$8.5\pm 1.7\times 10^2$&$34\pm 5$&$3.4\pm 1.8$&$2.3\pm 2.3\times 10^5$&$1.3\pm 1.3$&16\\
UMa I&$1.4\pm 0.4\times 10^4$&$318\pm 45$&$11.9\pm 3.5$&$2.6\pm 1.6\times 10^7$&$2.0\pm 1.2\times 10^{-1}$&3,8\\
UMa II&$4.0\pm 1.9\times 10^3$&$140\pm 25$&$6.7\pm 1.4$&$3.6\pm 1.6\times 10^6$&$3.2\pm 1.4\times 10^{-1}$&3,10\\
Willman 1&$1.0\pm 0.7\times 10^3$&$25\pm 6$&$4.3\pm 1.8$&$2.7\pm 2.3\times 10^5$&$4.1\pm 3.6$&3,8\\
\\
AndII&$9.3\pm 2.0\times 10^6$&$1230\pm 20$&$9.3\pm 2.7$&$6.2\pm 3.6\times 10^7$&$7.9\pm 4.5\times 10^{-3}$&17,18\\
AndIX&$1.8\pm 0.4\times 10^5$&$530\pm 110$&$6.8\pm 2.5$&$1.4\pm 1.1\times 10^7$&$2.3\pm 1.7\times 10^{-2}$&19\\
AndXV&$7.1\pm 1.4\times 10^5$&$270\pm 30$&$11\pm 6$&$1.9\pm 1.9\times 10^7$&$2.3\pm 2.3\times 10^{-1}$&20,21\\
Cetus&$2.8\pm 0.9\times 10^6$&$590\pm 20$&$17\pm 2$&$9.9\pm 2.3\times 10^7$&$1.1\pm 0.2\times 10^{-1}$&17,22\\
Sgr\tablenotemark{***}&$1.7\pm 0.3\times 10^7$&$1550\pm 50$&$11.4\pm 0.7$&$1.2\pm 0.1\times 10^8$&$7.5\pm 1.0\times 10^{-3}$&23,24\\
Tucana&$5.6\pm 1.6\times 10^5$&$270\pm 40$&$15.8\pm 3.6$&$4.0\pm 1.9\times 10^7$&$4.7\pm 2.3\times 10^{-1}$&25,26\\
\enddata
\tablenotetext{*}{Estimated using Equation 11}
\tablenotetext{**}{References: 1) \citet{ih95}; 2) \citet{walker09b}; 3) \citet{martin08}; 4) \citet{walker07b}; 5) \citet{mateo08}; 6) \citet{koch07b}; 7) Walker et al. in preparation; 8) \citet{martin07}; 9) \citet{koch09}; 10) \citet{simon07}; 11) \citet{aden09}; 12) \citet{belokurov08}; 13) \citet{walker09c}; 14) \citet{irwin07}; 15) \citet{geha09}; 16) \citet{belokurov09}; 17) \citet{mcconnachie06}; 18) \citet{cote99}; 19) \citet{chapman05}; 20) \citet{ibata07}; 21) \citet{letarte09}; 22) \citet{lewis07}; 23) \citet{ibata97}; 24) \citet{majewski03}; 25) \citet{saviane96}; 26) \citet{fraternali09}}
\tablenotetext{***}{Structural parameters refer to the bound central region of Sgr (see \citealt{majewski03}).}
\label{tab:dsphs}
\end{deluxetable*}

\subsection{Velocity Dispersion Profiles}

For the Jeans analysis in Section \ref{sec:jeans} we use empirical, projected velocity dispersion profiles for eight bright dSphs with abundant kinematic data.  \citet{walker07b} present velocity dispersion profiles for seven of these objects, but these profiles can now be updated---and we can now provide a profile for Ursa Minor---using the data we have obtained during the past two years.  For each galaxy we compute the velocity dispersion profile after 1) discarding all stars for which the probability of dSph membership, according to the algorithm described by \citet{walker09b}, is less than $0.95$; 2) subtracting the mild velocity gradients that likely reflect the bulk transverse motion of the dSph \citep{walker08}, and binning the data using circular annuli containing equal numbers of member stars.  We estimate the velocity dispersion in each bin using the maximum-likelihood technique described by \citet{walker06a}.  

\begin{figure*}
  \plotone{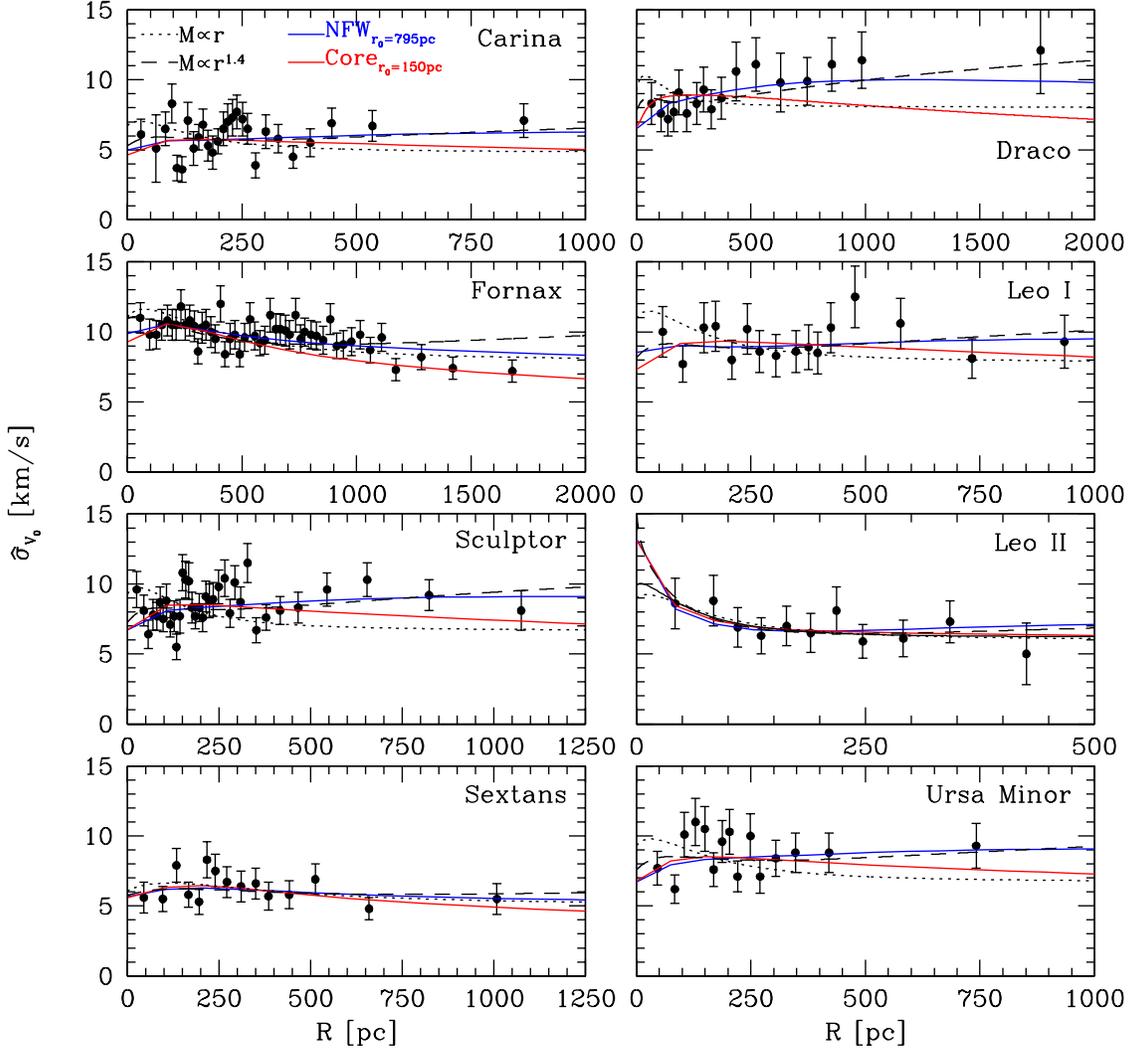}
  \caption{\scriptsize Projected velocity dispersion profiles for eight bright dSphs, from Magellan/MMFS and MMT/Hectochelle data.  Over-plotted are profiles calculated from isothermal, power-law, NFW and cored halos considered as prospective ``universal'' dSph halos (Section \ref{sec:universal}).  For each type of halo we fit only for the anisotropy and normalization.  All isothermal, NFW and cored profiles above have normalization $V_{max}\sim 10-20$ km s$^{-1}$---see Table \ref{tab:universal}.  All power-law profiles have normalization $M_{300}\sim [0.5-1.5]\times 10^7M_{\odot}$.}
  \label{fig:momentprofiles}
\end{figure*}

Figure \ref{fig:momentprofiles} displays the velocity dispersion profiles we measure for Carina, Draco, Fornax, Leo I, Leo II, Sculptor, Sextans and Ursa Minor.  While the profiles remain generally flat (c.f. Figure 2 of \citealt{walker07b}), we detect a gentle decline in velocity dispersion at large projected radii ($R \ga 1$ kpc) in Fornax.  This behavior becomes apparent only after addition of the most recent MMFS data \citep{walker09a}, which contribute most of the data points at large radius.  We also find hints of gently declining profiles in Sculptor and Sextans, although these data are noisier and the apparent declines hinge on the outermost one or two data points.  

Using the same data, \citet{lokas08} and \citet{lokas09} have recently measured profiles for Carina, Fornax, Leo I, Sculptor and Sextans that all decline more significantly at large radii than the profiles we present in Figure \ref{fig:momentprofiles}.  The difference arises because \citet{lokas08} and \citet{lokas09} adopt a rejection algorithm that attempts to remove stars that may have been tidally stripped and thus do not provide reliable tracers of the gravitational potential.  The algorithm, described in detail by \citet{klimentowski07} (see also \citealt{denhartog96}), iteratively estimates the mass profile from the velocities and positions of accepted members and then rejects stars with velocities that differ from the mean by more than $\sqrt{2GM(r)/r}$.  The validity of this method clearly depends on the accuracy of the mass profile estimate, and this quality is difficult to assess.  \citet{lokas08} and \citet{lokas09} use a mass estimator derived from the virial theorem \citep{heisler85}, which holds when the tracer population is self-gravitating or, if there is a dark matter component, when the mass-to-light ratio is constant with radius (``mass follows light'').  Then it is not surprising that after imposing this rejection algorithm \citet{lokas08} and \citet{lokas09} obtain falling velocity dispersion profiles that are consistent with mass-follows-light models.  While \citet{klimentowski07} succesfully test their algorithm and mass estimator using an N-body simulation of a tidally ``stirred'' \citep{mayer01a,mayer01b} satellite in which mass actually does follow light (after 99\% of the initial dark mass is stripped), they do not demonstrate that their method preserves flat velocity dispersion profiles where they may genuinely exist---e.g., in  equilibrium systems with extended and intact dark matter halos\footnote{\citet{klimentowski07} note that their method rejects fewer than 1\% of stars drawn from an equilibrium system with an extended dark matter halo, but they do not describe the locations of the rejected stars or the impact on the measured velocity dispersion profile.}.  

Pending evidence that the tidally stirred N-body model of \citet{klimentowski07} describes the evolution of the real dSphs in our sample, we opt not to apply the rejection method used by \citet{lokas08,lokas09}.  Thus we avoid imposing any bias toward mass-follows-light models that may result from use of the virial theorem mass estimator, but at the cost of leaving our samples vulnerable to contamination from tidally unbound stars.  Given the difficulty of identifying such stars, we choose to err on the side of including them in our analysis.  For at least some systems---e.g., Fornax, which has orbital pericenter near its present distance of $\sim 140$ kpc \citep{piatek07,walker08}---we expect a negligible contribution from unbound tidal debris.  Among the remaining dSphs in our sample, there exists kinematic evidence of tidal streaming motion in the outer parts of Carina \citep{munoz06} and Leo I \citep{sohn07,mateo08}.  In fitting models to the velocity dispersion profiles of these two galaxies, as well as Draco, which shows a rising profile in its outer parts, we experimented with discarding the outermost 1-2 data points in each system.  Since the results were indistinguishable from fits that included the outermost points, we decided to include the outer data points in our final analysis.  

\section{Jeans Analysis} 
\label{sec:jeans}

With the goal of measuring dSph masses, we assume that the data sample in each dSph a single, pressure-supported stellar population that is in dynamic equilibrium and traces an underlying gravitational potential dominated by dark matter.  Further assuming spherical symmetry, the mass profile, $M(r)$, of the dark matter halo relates to (moments of) the stellar distribution function via the Jeans equation:
\begin{equation}
\frac{1}{\nu}\frac{d}{dr}(\nu \bar{v_r^2})+2\frac{\beta\bar{v_r^2}}{r}=-\frac{GM(r)}{r^2},
\label{eq:jeans}
\end{equation}
where $\nu(r)$, $\bar{v_r^2}(r)$, and $\beta(r)\equiv 1-\bar{v_{\theta}^2}/\bar{v_r^2}$ describe the 3-dimensional density, radial velocity dispersion, and orbital anisotropy, respectively, of the stellar component.  For the special case of constant anisotropy, the Jeans equation has the solution (e.g., \citealt{mamon05})
\begin{equation}
  \nu\bar{v^2_r}=Gr^{-2\beta}\displaystyle\int_r^{\infty}s^{2\beta-2}\nu(s)M(s)ds.
  \label{eq:jeanssolution}
\end{equation}
Projecting along the line of sight, the mass profile relates to observable profiles, the projected stellar density, $I(R)$, and velocity dispersion, $\sigma_p(R)$, according to \citep[``BT08'' hereafter]{bt08}
\begin{equation}
  \sigma_p^2(R)=\frac{2}{I(R)}\displaystyle \int_{R}^{\infty}\biggl (1-\beta\frac{R^2}{r^2}\biggr ) \frac{\nu \bar{v_r^2}r}{\sqrt{r^2-R^2}}dr.
  \label{eq:jeansproject}
\end{equation}
To estimate dSph masses via the Jeans equation we therefore employ the following strategy: 1) adopt a simple analytic profile for $I(R)$ from the literature; 2) adopt a parametric model for $M(r)$; and 3) find the halo parameters that, via equations \ref{eq:jeanssolution} and \ref{eq:jeansproject}, best reproduce the empirical velocity dispersion profiles shown in Figure \ref{fig:momentprofiles}.

\subsection{Stellar Density}
\label{subsec:stellardensity}

Stellar surface densities of dSphs are typically fit by \citet{plummer11}, \citet{king62} and/or \citet{sersic68}, profiles (e.g., \citealt{ih95,mcconnachie06,belokurov07}).  The Plummer profile, $I(R)=L(\pi r_{half}^2)^{-1}[1+R^2/r_{half}^2]^{-2}$ where $L$ is the total luminosity, is the simplest as it has only a single shape parameter, the projected half-light radius\footnote{Throughout this work we define $r_{half}$ as the two-dimensional half-light radius---i.e., the radius of the cylinder that encloses half of the total luminosity.}.  It is also the only profile with published parameters for all dSphs, since the concentration parameters of King and Sersic profiles are not well-constrained by the sparse data available for the faintest dSphs.  Therefore, in what follows we adopt the Plummer profile to characterize dSph stellar densities.  

Given a model $I(R)$ for the projected stellar density, one recovers the 3-dimensional density from (BT08)
\begin{equation}
  \nu(r)=-\frac{1}{\pi}\displaystyle\int_r^{\infty}\frac{dI}{dR}\frac{dR}{\sqrt{R^2-r^2}}.
  \label{eq:deproject}
\end{equation}
Thus for the Plummer profile, we have $\nu(r)=3L(4\pi r_{half}^3)^{-1}[1+r^2/r_{half}^2]^{-5/2}$.

We note that even though dSph surface brightness data can be fit adequately by a variety of density profiles, the choice of profile is not trivial.  \citet{evans09} demonstrate that, even when the gravitational potential is dominated by dark matter, the adopted shape of the stellar density profile can profoundly affect the inferred shape of $M(r)$ at small radii.  In what follows, while for simplicity we present only the results obtained using the Plummer profile, we explicitly identify any results that are strongly sensitive to this choice (Section \ref{subsubsec:corecusp}).

\subsection{Halo Model}
\label{subsec:halomodel}

For the dark matter halo we follow S08 (also \citealt{koch07,koch07b}) in adopting a generalized Hernquist profile given by \citep{hernquist90,zhao96}
\begin{equation}
  \rho(r)=\rho_0\biggl (\frac{r}{r_0}\biggr )^{-\gamma}\biggl [1+\biggl (\frac{r}{r_0}\biggr )^{\alpha}\biggr ]^{\frac{\gamma-3}{\alpha}},
  \label{eq:hernquist1}
\end{equation}
where the parameter $\alpha$ controls the sharpness of the transition from inner slope $\lim_{r\rightarrow 0}d\ln(\rho)/d\ln(r)\propto -\gamma$ to outer slope $\lim_{r\rightarrow \infty}d\ln(\rho)/d\ln(r)\propto -3$.  In terms of these parameters, the mass profile is
\begin{eqnarray}
  M(r)=4\pi\displaystyle\int_{0}^{r}s^2\rho(s)ds=\frac{4\pi \rho_0r_0^3}{3-\gamma}\biggl (\frac{r}{r_0}\biggr )^{3-\gamma}\hspace{0.6in}\\
  _2F_1\biggl [\frac{3-\gamma}{\alpha},\frac{3-\gamma}{\alpha};\frac{3-\gamma+\alpha}{\alpha};-\biggl (\frac{r}{r_0}\biggr )^\alpha\biggr ],\nonumber
  \label{eq:hernquist2}
\end{eqnarray}
where $_2F_1(a,b;c;z)$ is Gauss's hypergeometric function.  

The profiles admitted by Equation \ref{eq:hernquist1} include the range of plausible halo shapes relevant to the ongoing controversy regarding ``cores'' versus ``cusps'' in individual dark matter halos.  Profiles with $\gamma>0$ are centrally cusped while those with $\gamma=0$ have constant-density cores.  For $\alpha=\gamma=1$, one recovers the cuspy \citet[``NFW'' hereafter]{navarro96,navarro97} profile motivated by cosmological N-body simulations.  

\subsection{Normalization}
\label{subsec:normalization}

It is convenient to normalize the mass profile by $M(r_0)$, the enclosed mass at the scale radius of the dark matter halo.  This quantity relates to the maximum circular velocity, $V_{max}$, according to
\begin{equation}
  M(r_0)=\frac{\eta r_0V_{max}^2}{G}\frac{_2F_1\bigl [\frac{3-\gamma}{\alpha},\frac{3-\gamma}{\alpha};\frac{3-\gamma+\alpha}{\alpha};-1\bigr ]}{\eta^{3-\gamma}\hspace{0.001cm}_2F_1\bigl [\frac{3-\gamma}{\alpha},\frac{3-\gamma}{\alpha};\frac{3-\gamma+\alpha}{\alpha};-\eta^\alpha\bigr ]},
  \label{eq:mr0}
\end{equation}
where $\eta\equiv r_{max}/r_0$ identifies the radius corresponding to the maximum circular velocity, and is specified uniquely by $\alpha$ and $\gamma$.  Thus the parameter $V_{max}$ sets the normalization of the mass profile.  

The normalization can equivalently be set by specifying, rather than $V_{max}$, the enclosed mass at some particular radius.  For radius $x$, the enclosed mass $M(x)$ specifies $M(r_0)$ according to 
\begin{equation}
  M(r_0)=M(x)\frac{_2F_1\bigl [\frac{3-\gamma}{\alpha},\frac{3-\gamma}{\alpha};\frac{3-\gamma+\alpha}{\alpha};-1\bigr ]}{\bigl (\frac{x}{r_0}\bigr )^{3-\gamma}\hspace{0.001cm}_2F_1\bigl [\frac{3-\gamma}{\alpha},\frac{3-\gamma}{\alpha};\frac{3-\gamma+\alpha}{\alpha};-\bigl (\frac{x}{r_0}\bigr )^{\alpha}\bigr ]}.
  \label{eq:mx}
\end{equation}
S08 demonstrate that for most dSphs the Jeans analysis can tightly constrain $M_{300}$.  Here, in addition to $M_{300}$, we shall consider the masses within two alternative radii as free parameters with which to normalize the mass profile.  Specifically, we consider the mass within the half-light radius, $M(r_{half})$, and the mass within the outermost data point of the empirical velocity dispersion profile, $M(r_{last})$.  

\subsection{Markov-Chain Monte Carlo Method}
\label{subsec:mcmc}

In order to evaluate a given halo model, we compare the projected velocity dispersion profile, $\sigma_p(R)$, from Equation \ref{eq:jeansproject} to the empirical profile, $\sigma_{V_0}(R)$, displayed in Figure \ref{fig:momentprofiles}.  For a given parameter set $S\equiv \{-\log (1-\beta),\log M_X,\log r_0,\alpha,\gamma\}$, where $M_X$ is one of $\{V_{max}$, $M(r_{half})$, $M_{300}$ or $M(r_{last})\}$, we adopt uniform priors and consider the likelihood
\begin{equation}
  \zeta=\displaystyle \prod_{i=1}^N \frac{1}{\sqrt{2\pi(\mathrm{Var}[\sigma_{V_0}(R_i)])}}\exp\biggl [-\frac{1}{2}\frac{(\sigma_{V_0}(R_i)-\sigma_p(R_i))^2}{\mathrm{Var}[\sigma_{V_0}(R_i)]}\biggr ],
  \label{eq:likelihood}
\end{equation}
where $\mathrm{Var}[\sigma_{V_0}(R_i)]$ is the square of the error associated with the empirical dispersion.  

Our mass models have five free parameters (four halo parameters plus one anisotropy parameter).  In order to explore the large parameter space efficiently, we employ Markov-chain Monte Carlo (MCMC) techniques.  That is, we use the standard Metropolis-Hastings algorithm \citep{metropolis53,hastings70} to generate posterior distributions according to the following prescription: 1) from the current location in parameter space, $S_n$, draw a prospective new location, $S'$, from a Gaussian probability density centered on $S_n$; 2) evaluate the ratio of likelihoods at $S_n$ and $S'$; and 3) if $\zeta(S')/\zeta(S_n)\ge 1$, accept such that $S_{n+1}=S'$, else accept with probability $\zeta(S')/\zeta(S_n)$, such that $S_{n+1}=S'$ with probability $\zeta(S')/\zeta(S_n)$ and $S_{n+1}=S_n$ with probability $1-\zeta(S')/\zeta(S_n)$.  

For this procedure we use the adaptive MCMC engine CosmoMC\footnote{available at http://cosmologist.info/cosmomc} \citep{lewis02}.  Although it was developed specifically for analysis of cosmic microwave background data, CosmoMC provides a generic sampler that continually updates the probability density according to the parameter covariances in order to optimize the acceptance rate.  For each galaxy and parameterization we run four chains simultaneously, allowing each to proceed until the variances of parameter values across the four chains become less than 1\% of the mean of the variances.  Satisfaction of this convergence criterion typically requires $\sim 10^4$ steps for our chains.  We then identify the posterior distribution in parameter space with the last 70\% of all accepted points (we discard the first 30\% of points, which correspond to the ``burn-in'' period.)

\subsection{Results}
\label{subsec:results}

\begin{figure*}
  \plotone{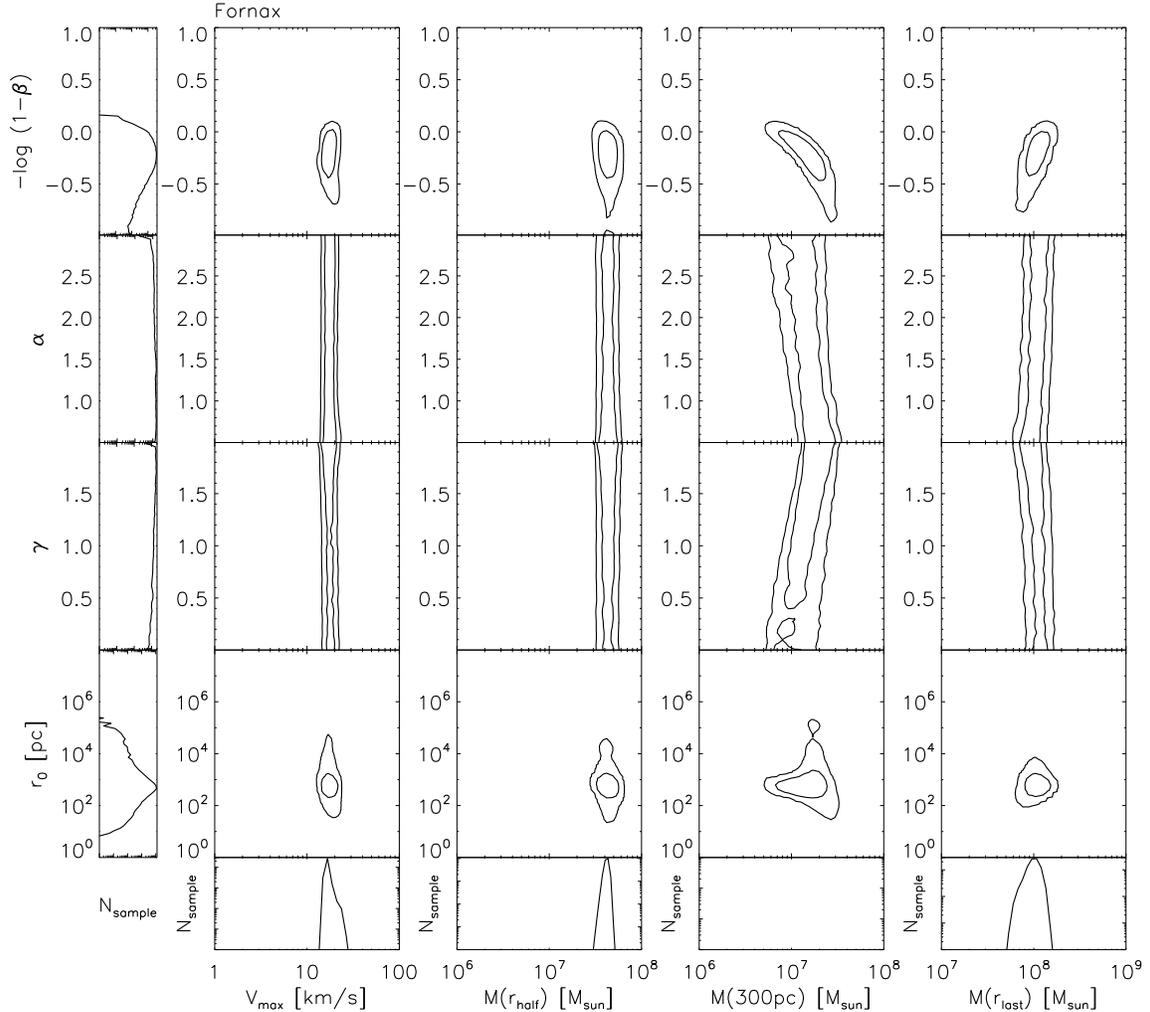}
  \caption{\scriptsize Two-dimensional posterior distributions of Fornax halo model parameters from Markov-Chain Monte Carlo (MCMC) analysis (section \ref{sec:jeans}).  Contours identify regions of 68\% and 95\% confidence, respectively.  Histograms indicate marginal distributions of each parameter.   }
  \label{fig:for_par5}
\end{figure*}

Figure \ref{fig:for_par5} indicates posterior distributions of model parameters that we obtain from our MCMC analysis of Fornax, for each choice of parameter used to specify the normalization of the mass profile.  We ran CosmoMC independently for each normalization parameter, so while the distributions for other parameters should be similar for a given galaxy, they need not be identical.  Contours in Figure \ref{fig:for_par5} identify regions of 68\% and 95\% confidence.  

In general there exist degeneracies among the five parameters such that most cannot be constrained uniquely.  However, we find that most allowed mass profiles, regardless of their shapes, intersect near the half-light radius.  Then, as Figure \ref{fig:for_par5} demonstrates, when we normalize the mass profile by either $M(r_{half})$ or $M_{300}$, the degeneracies are such that the allowed region in parameter space subtends a relatively small mass range. Therefore while our analysis cannot place meaningful constraints on most parameters, it does constrain $M(r_{half})$ and $M_{300}$.  At much smaller and much larger radii (e.g., $r_{last}$), the allowed mass profiles can diverge and the resulting constraints are weaker.  

\begin{figure*}
  \plotone{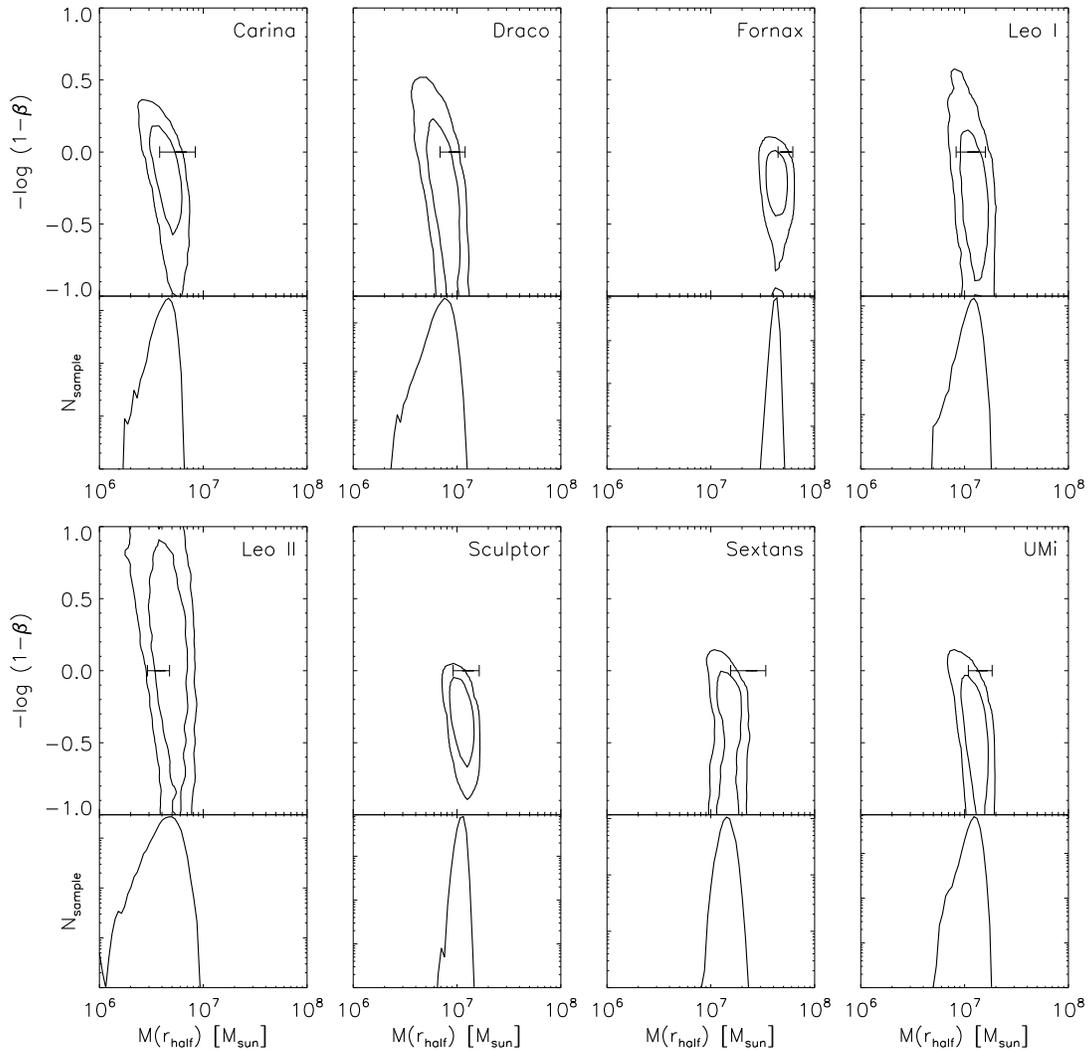}
  \caption{\scriptsize Posterior distributions of halo parameters specifying anisotropy and $M(r_{half}$, from Markov-Chain Monte Carlo (MCMC) analysis (section \ref{sec:jeans}), for each of the classical dSphs.  Contours identify regions of 68\% and 95\% confidence, respectively.  Histograms indicate marginal distributions of $M(r_{half})$.  For comparison, error bars indicate estimates obtained using Equation \ref{eq:mrhalf}.}
  \label{fig:all_mrhalfbeta}
\end{figure*}

In the interest of brevity we do not include the equivalent of Figure \ref{fig:for_par5} for each of the other galaxies in our sample.  Instead, for each of the eight dSphs with velocity dispersion profiles in Figure \ref{fig:momentprofiles}, Figure \ref{fig:all_mrhalfbeta} indicates the posterior distributions of $M(r_{half})$ and anisotropy.  Table \ref{tab:mcmc} then lists the constraints on $M(r_{half})$, $M_{300}$, $M(r_{last})$ and $V_{max}$.  

\begin{deluxetable*}{lrrrrrrrrrr}
\tabletypesize{\scriptsize}
\tablewidth{0pc}
\tablecaption{Mass constraints from MCMC analysis and Eq. 10}
\tablehead{\\
  \colhead{Galaxy}&\colhead{Method}&\colhead{$M(r_{\mathrm{half}})$}&\colhead{$M_{300}$}&\colhead{$r_{last}$; $M(r_{last})$}&\colhead{$V_{max}$}\\
  \colhead{}&\colhead{}&\colhead{$[10^7M_{\odot}]$}&\colhead{$[10^7M_{\odot}]$}&\colhead{[kpc]; $[10^7M_{\odot}]$}&\colhead{[km s$^{-1}$]}\\
}
\startdata
Carina&MCMC&$0.4_{-0.1}^{+0.1}$&$0.6_{-0.2}^{+0.1}$&$0.87; 3.7_{-1.8}^{+2.1}$&$\ge 10$\\
Carina&Eq. 10&$0.6\pm 0.2$&$0.9\pm 0.3$&\nodata&\nodata\\
\\
Draco&MCMC&$0.6_{-0.3}^{+0.5}$&$1.5_{-0.6}^{+0.4}$&$0.92; 26.4_{-17.4}^{+18.6}$&$\ge 17$\\
Draco&Eq. 10&$0.9\pm 0.3$&$2.0\pm 0.5$&\nodata&\nodata\\
\\
Fornax&MCMC&$4.3_{-0.7}^{+0.6}$&$0.7_{-0.2}^{+2.0}$&$1.7; 12.8_{-5.6}^{+2.2}$&$18_{-3}^{+5}$\\
Fornax&Eq. 10&$5.3\pm 0.9$&$0.8\pm 0.1$&\nodata&\nodata\\
\\
Leo I&MCMC&$1.0_{-0.4}^{+0.6}$&$1.6_{-0.4}^{+0.4}$&$0.93; 8.9_{-5.2}^{+4.3}$&$\ge 15$\\
Leo I&Eq. 10&$1.2\pm 0.4$&$1.8\pm 0.5$&\nodata&\nodata\\
\\
Leo II&MCMC&$0.5_{-0.3}^{+0.2}$&$1.3_{-0.7}^{+0.7}$&$0.42; 1.7_{-1.2}^{+1.9}$&$\ge 11$\\
Leo II&Eq. 10&$0.4\pm 0.1$&$1.2\pm 0.3$&\nodata&\nodata\\
\\
Sculptor&MCMC&$1.0_{-0.3}^{+0.3}$&$1.3_{-0.2}^{+0.2}$&$1.1; 10.0_{-5.0}^{+3.2}$&$\ge 15$\\
Sculptor&Eq. 10&$1.3\pm 0.4$&$1.7\pm 0.5$&\nodata&\nodata\\
\\
Sextans&MCMC&$1.6_{-0.4}^{+0.4}$&$0.7_{-0.4}^{+0.6}$&$1.0; 2.0_{-0.7}^{+1.0}$&$\ge 9$\\
Sextans&Eq. 10&$2.5\pm 0.9$&$0.4\pm 0.2$&\nodata&\nodata\\
\\
Ursa Minor&MCMC&$1.3_{-0.5}^{+0.3}$&$1.4_{-0.4}^{+0.3}$&$0.74; 4.4_{-2.0}^{+2.9}$&$\ge 13$\\
Ursa Minor&Eq. 10&$1.5\pm 0.4$&$1.7\pm 0.4$&\nodata&\nodata\\
\enddata
\label{tab:mcmc}
\end{deluxetable*}

\subsubsection{Mass}
\label{subsubsec:constraints}

We confirm the S08 result that the Jeans/MCMC analysis places relatively tight constraints on $M_{300}$, and that for the eight galaxies considered here, $M_{300}\sim 10^7M_{\odot}$.  We obtain similarly tight constraints on $M(r_{half})$, which is not surprising, since for these galaxies $r_{half}\sim 300$ pc.  These results are not sensitive to the adopted form of the stellar density profile---we obtain similar values when we repeat the analysis using exponential and King profiles.  Table \ref{tab:mcmc} lists the masses and 68\% confidence intervals obtained from the MCMC analysis.

\subsubsection{$V_{max}$ for Fornax}
\label{subsubsec:vmax}

Using similar methods with smaller data sets, \citet{strigari06} and \citet{penarrubia08a} demonstrate a degeneracy between parameters $V_{max}$ and $r_0$ such that halos with arbitrarily large $V_{max}$ can, given the freedom to make $r_0$ also arbitrarily large, all have similar masses within the luminous region.  However, whereas most previously published velocity dispersion profiles are quite flat\footnote{\citet{wilkinson04} and \citet{kleyna04} report \textit{sharp} declines at large radii in Ursa Minor and Sextans, but these features are not reproduced either in our data sets or in those of \citet{munoz05}.  In any case such a sudden decline is difficult to reconcile with equilibrium models.} (e.g., \citealt{kleyna02,koch07,koch07b,walker07b}), we now detect gently decreasing dispersion in the outer regions of Fornax (Figure \ref{fig:momentprofiles}).  In our Jeans/MCMC analysis this behavior constrains the scale radius of Fornax's dark matter halo.  Although the constraint is loose, $r_0\sim 0.49_{-0.45}^{+9}$ kpc, it is enough to help break the degeneracy with $V_{max}$; in tests during which we artificially boost the velocity dispersion in the outer bins such that Fornax's dispersion profile would be perfectly flat, our analysis fails to place upper limits on either the scale radius or $V_{max}$.  Thus with the addition of the outer Fornax data, we now obtain the first model-independent constraint on $V_{max}$ for any dSph galaxy: $V_{max}=18_{-3}^{+5}$ km s$^{-1}$.  For the remaining dSphs, the absence of any constraint on $r_0$ implies that we can place only lower limits on $V_{max}$, typically of $V_{max}\ga 10$ km s$^{-1}$ (see Table \ref{tab:mcmc}).   

One caveat regarding the apparent constraints on Fornax's $V_{max}$ and $r_0$ follows from the fact that we have assumed the stellar velocity anisotropy, $\beta$, is constant.  Our models therefore associate the onset of declining velocity dispersion with a transition in the shape of the density profile, which in our models helps to specify $r_0$.  In reality, falling dispersions may alternatively signal the onset of significant anisotropy.  Therefore any constraints on $r_0$, and in turn on $V_{max}$, must be interpreted with some caution.  However, the same is not true of the bulk mass constraints---e.g., S08's models allow for radially variable anisotropy and demonstrate that constraints on $M(r_{half})$ and $M_{300}$ are robust over the full range of models considered.

\subsubsection{No Constraint on Cores/Cusps}
\label{subsubsec:corecusp}

For no dSph does our Jeans analysis place meaningful constraints on the halo shape parameters $\alpha$ and $\gamma$.  For the reasons explained by \citet{evans09}, any such constraint on the inner density slope, $\gamma$, is strongly sensitive to the shape of the adopted stellar density model\footnote{For a simple illustration, consider the mass profile given by the first part of Equation \ref{eq:simplejeansmass}, which is derived from the Jeans equation in the special case of $\beta=0$ and $\bar{v_r^2}=$const.  Then the shape of $M(r)$ is determined uniquely by the stellar density profile.}.  Meaningful constraints on $\gamma$ require more sophisticated analyses that incorporate realistic models of the stellar distribution function (e.g., \citealt{wilkinson02}; Wilkinson et al. in prep.) and can thereby be evaluated on a star-by-star basis. 

\subsection{A Simple and Robust Estimator for $M(r_{half})$}
\label{subsec:simple}

We conclude that the available data tightly constrain dSph masses within $r\sim r_{half}$.  Furthermore, these constraints hold over a wide range of possible halo shapes and stellar velocity anisotropies.  S08 reach the same conclusion regarding the robustness of $M_{300}$, while \citet{penarrubia08a} show that $M(r_{half})$ is well constrained for an NFW halo regardless of the particular values of $V_{max}$ and $r_0$.  Our analysis thus confirms the S08 result and generalizes the conclusion of \citet{penarrubia08a} to include non-NFW halos.  In short: we know $M(r_{half})$ for the dwarf spheroidals.  

Because the MCMC method requires significant computational effort, it is reasonable to wonder whether the measurement of such a bulk quantity as $M(r_{half})$ might adequately be made using more conventional techniques.  Therefore let us define a simple analytic model that relates $M(r_{half})$ to observed quantities.  Specifically, suppose the stellar component is distributed as a Plummer sphere with a velocity distribution that is isotropic ($\beta=0$) and has constant dispersion $\sigma_{V_0}^2(R)=\bar{v_r^2}=\sigma^2$.  Since these conditions are broadly consistent with all available dSph data\footnote{Velocity dispersion profiles are not yet available for the faintest dSphs, for which small samples provide measurements only of central velocity dispersions.  Thus our assumption that the faintest dSphs have flat velocity dispersion profiles remains untested.}, this exercise amounts to consideration of a subset of the models not ruled out by the MCMC analysis discussed above.  Then the Jeans equation gives
\begin{equation}
M(r)=-\frac{r^2\bar{v_{r}^2}}{G\nu}\frac{d\nu}{dr}=\frac{5r_{half}\sigma^2\bigl (\frac{r}{r_{half}}\bigr )^3}{G\bigl [1+r^2/r_{half}^2\bigr ]},
\label{eq:simplejeansmass}
\end{equation}
where the last expression is specific to the assumption of a Plummer profile for the stellar density.  Equation \ref{eq:simplejeansmass} immediately provides the convenient estimate 
\begin{equation}
  M(r_{half})=\mu r_{half}\sigma^2,
  \label{eq:mrhalf}
\end{equation}
where $\mu\equiv 580$ $M_{\odot}$pc$^{-1}$km$^{-2}$s$^2$, which differs only by a factor of $1.5$ from the core-fitting formula commonly used to estimate dSph dynamical masses (e.g., \citealt{illingworth76,richstone86,mateo98}).  

\begin{figure}
  \epsscale{1.2}
  \plotone{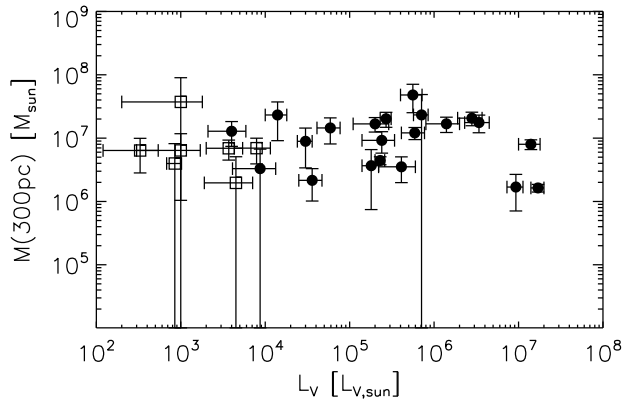}
  \caption{\scriptsize $M_{300}$, estimated using Equation \ref{eq:simplejeansmass}, versus total V-band luminosity.  The flat relation reproduces the main result of \citet{strigari08} (see their Figure 1).  Filled circles represent galaxies for which observed kinematic tracers extend beyond $r\ga 300$ pc.  Open squares represent ultra-faint satellites for which $r_{half}< 100$ pc, and for which there is no available evidence that a dark matter halo actually extends to a radius of 300 pc.}
  \label{fig:m300}
\end{figure}

Table \ref{tab:mcmc} lists estimates of $M(r_{half})$ and $M_{300}$ obtained from Equations \ref{eq:simplejeansmass}-\ref{eq:mrhalf} for the eight bright dSphs with velocity dispersion profiles, and Figure \ref{fig:all_mrhalfbeta} indicates the estimates obtained from Equation \ref{eq:mrhalf} for easy comparison to estimates obtained from the full Jeans/MCMC analysis.  We find that the simple estimates from Equation \ref{eq:mrhalf} generally stand in excellent agreement with constraints we obtain from the full Jeans/MCMC analysis.  Furthermore, as shown in Figure \ref{fig:m300}, we reproduce the flat $M_{300}$-luminosity relation of S08 (c.f. their Figure 1) accurately merely by applying Equation \ref{eq:simplejeansmass} to the data listed in Table \ref{tab:dsphs}.  We conclude that the mass estimates given by Equations \ref{eq:simplejeansmass}-\ref{eq:mrhalf} are robust against a wide range of halo models when evaluated near the half-light radius.  At radii much different from the half-light radius, the estimates become model-dependent.  In what follows, the values of $M(r_{half})$ that we consider are those obtained from Equation \ref{eq:mrhalf}.

\section{A New Scaling Relation for dSphs}
\label{sec:scaling}

\begin{figure*}
  \epsscale{1.2}
  \plotone{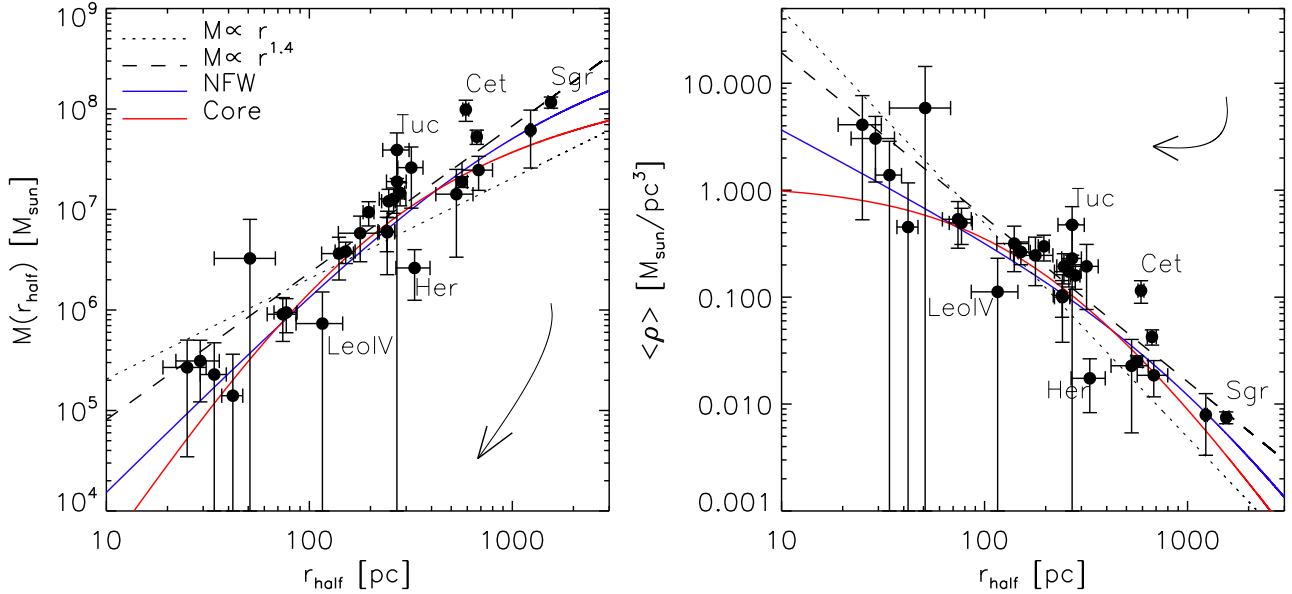}
  \caption{\scriptsize \textit{Left:} $M(r_{half})$ against half-light radius for all measured dSphs and dSph candidates (see Table \ref{tab:dsphs}).  Over-plotted are the best-fitting mass profiles, $M(r)$, for isothermal, power-law, NFW ($V_{max}=15$ km s$^{-1}$, $r_0=795$ pc) and cored ($V_{max}=13$ km s$^{-1}$, $r_0=150$ pc) halo profiles.  \textit{Right:} Mean density within the half-light radius, versus $r_{half}$.  Over-plotted are the same halo models as in the left panel.  In both panels, arrows indicate the magnitude and direction that individual galaxies would be displaced due to the tidal stripping of 99\% of the original stellar mass \citep{penarrubia08b}.  Text markers identify the most extreme outliers and Sagittarius.}
  \label{fig:mflat2}
\end{figure*}

\begin{figure}
  \epsscale{1.2}
  \plotone{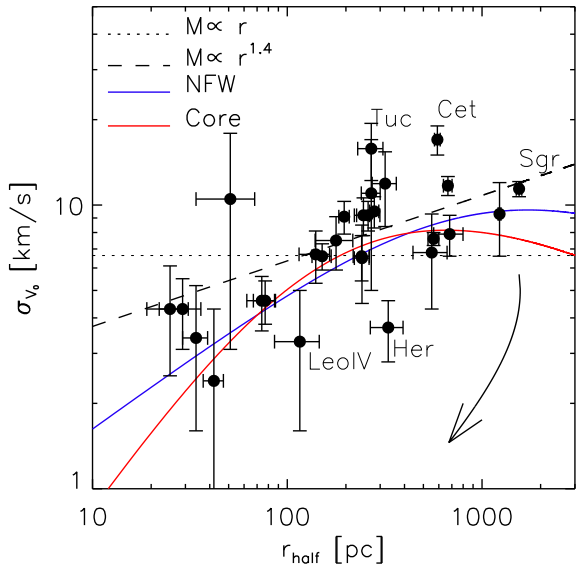}
  \caption{\scriptsize Global velocity dispersion against half-light radius for all measured dSphs and dSph candidates (see Table \ref{tab:dsphs}).  Over-plotted are the scaling relations that correspond to the best-fitting mass profiles shown in Figure \ref{fig:mflat2}.  For the NFW and cored halos, these are translated into the $\sigma$-$r_{half}$ plane via Equations \ref{eq:nfwsigma}-\ref{eq:coresigma}.  As in Figure \ref{fig:mflat2}, the arrow indicates the magnitude and direction of displacement due to the tidal stripping of 99\% of the original stellar mass \citep{penarrubia08b}.}
  \label{fig:mflat3}
\end{figure}

The left-hand panel of Figure \ref{fig:mflat2} plots $M(r_{half})$ (from Equation \ref{eq:mrhalf}) against $r_{half}$ for the 28 objects with published kinematic data (see Table \ref{tab:dsphs}).  There we find an approximately power-law correlation between $M(r_{half})$ and $r_{half}$.  Of course, one expects power-law behavior merely from the form of Equation \ref{eq:mrhalf}---e.g., in the absence of a correlation between half-light radius and velocity dispersion, one expects to fit the mass profile of a singular isothermal sphere, $M(r)\propto r$ (dotted line in Figure \ref{fig:mflat2}).  Yet the slope in the empirical $M(r_{half})-r_{half}$ relation is steeper than that corresponding to the isothermal sphere.  This result follows entirely from the fact that, given the data now available for ultra-faint dSphs, there exists a clear correlation, shown in Figure \ref{fig:mflat3}, between dSph velocity dispersion and half-light radius.  Fitting a power law in the plane of directly observed quantities, we find $\log [\sigma_{V_0}/(\mathrm{km s}^{-1})]\sim 0.2\log [r_{half}/\mathrm{pc}]+0.3$ (long-dashed line in Figure \ref{fig:mflat3}).  Using Equation \ref{eq:mrhalf} to translate into the mass-radius plane, we obtain $M(r_{half})/M_{\odot}\sim 2300 (r_{half}/\mathrm{pc})^{1.4\pm 0.4}$ (long-dashed line in Figure \ref{fig:mflat2}).


Constraints on $M(r_{half})$ translate directly to give the mean density interior to the half-light radius, $\langle \rho\rangle\equiv 3 M(r_{half})/(4\pi r_{half}^3)$.  The right-hand panel of Figure \ref{fig:mflat2} plots mean density against half-light radius for the entire dSph sample.  In terms of mean density inside $r_{half}$, the aforementioned relations imply $\langle \rho\rangle /(M_{\odot}\mathrm{pc}^{-3})\sim 550(r_{half}/\mathrm{pc})^{-1.6\pm 0.4}$, and indeed, the data points in the right panel of Figure \ref{fig:mflat2} are well fit by this relation (long-dashed line).  We note in particular that the dSph population adheres to this relation even as $r_{half}$ reaches values as small as tens of parsecs, with the caveat that in some cases the measured velocity dispersions, densities and masses are upper limits.  If their velocity dispersions have indeed been resolved and reliably trace their masses, the smallest dSphs exhibit the largest galactic densities yet recorded, with $\langle \rho \rangle \la 5 M_{\odot}$pc$^{-3}\sim 200$ GeV cm$^{-3}$.  These extreme values point to an early epoch of formation, when the Universe itself would have had similar density, and may account for the survival of the ultra-faint systems against the destructive influence of tides (see below).  

\subsection{The Effects of Tides}
\label{subsec:tides}

The observable properties of the Local Group dSphs must be influenced to some degree by the tidal forces exerted by the Milky Way and M31.  Early work on this problem (e.g., \citealt{oh95,pp95}) suggested that tides do not significantly alter the central velocity dispersion of a satellite as it passes through pericenter, lending credibility to estimates of dSph dynamical masses.  More recent N-body simulations (e.g., \citealt{read06,munoz08,penarrubia08b,klimentowski09}) monitor the evolution of both stellar and dark matter components over several pericentric passages and a wide range of orbits.  \citet[``P08'' hereafter]{penarrubia08b} provide a set of analytic formulae to describe the tidal evolution of the parameters relevant to the present work.  In particular, they find that for a dSph consisting of a stellar component described by a King profile embedded within an NFW halo, all parameters evolve according to the fraction of mass lost within the core radius (here taken to be approximately the half-light radius).  For tidal stripping that results in the loss of as much as 99\% of the original luminosity, the evolution of parameters is described by $L/L_0\approx 2^7(\sigma/\sigma_0)^{6.6}[1+\sigma^2/\sigma_0^2]^{-7}$ and $r_{half}/r_{half_0}\approx 2^{3/2}(\sigma/\sigma_0)^{1.3}[1+\sigma^2/\sigma_0^2]^{-3/2}$, where subscripts of naught indicate initial values (P08).  

Arrows in Figures \ref{fig:mflat2} and \ref{fig:mflat3} indicate the directions and magnitudes of the tidal tracks followed by a satellite as it loses 99\% of its stellar mass under the P08 relations.  In the $\sigma$-$r_{half}$ plane of obvervables (Figure \ref{fig:mflat3}), at all points along the tidal track the displacement vector points primarily downward such that, as the satellite loses mass, its velocity dispersion decreases more quickly than its size.  As a result, tides tend to carry individual objects off the empirical relation rather than move them along it.  While the tidal track in the $M(r_{half})$-$r_{half}$ plane (Figure \ref{fig:mflat2}, left) is more parallel to the empirical relation, the tidal track in the $\langle \rho\rangle$-$r_{half}$ plane (right panel of Figure \ref{fig:mflat2}) provides strong evidence against a tidal origin for the empirical relation.  There we find that tides tend to decrease both the half-light radius \textit{and} the mean density within the half-light radius.  Thus tides do not generate the circumstance in which the smallest objects also have the largest mean densities interior to $r_{half}$.  These considerations lead us to conclude that tides did not introduce the correlations exhibited in Figures \ref{fig:mflat2} and \ref{fig:mflat3}.  This is not to say that tides are irrelevant---we now discuss two ways in which tides may have altered the character of the correlations we now observe.  

First, we can expect at least some of the scatter in the correlations to be due to the varying degrees to which dSphs on different orbits have been tidally sculpted.  In this regard, the dSphs Cetus and Tucana merit specific attention because they are the only objects in our sample that have no clear association with either the Milky Way or M31.  Cetus lies at a distance of $D\sim 755$ kpc \citep{mcconnachie06} and Tucana is even more remote, at $D\sim 880$ kpc \citep{saviane96}.  It is therefore reasonable to assume that neither object has lost appreciable amounts of mass to tidal stripping.  Then it is perhaps not a coincidence that Cetus and Tucana are the two most extreme outliers with respect to the empirical relations in Figures \ref{fig:mflat2} and \ref{fig:mflat3}.  Both objects have large velocity dispersions (and hence large $M(r_{half})$) compared to other objects of similar size.  We speculate that tides may have influenced \textit{all} other dSphs to a significantly higher degree than Cetus and Tucana, moving the population systematically along P08's tidal tracks while leaving Cetus and Tucana in relatively pristine condition.  Notice that even strong tidal stripping, if it acts to a similar extent on most dSphs, can preserve the slope of an initial correlation despite shifting the locus of the affected population.

Second, tides might alter the slopes of the correlations in Figures \ref{fig:mflat2} and \ref{fig:mflat3} if systems of a particular size happen to be the most prone to tidal stripping.  According to the tidal tracks, tides act to move objects primarily downward, towards lower velocity dispersion in the $\sigma$-$r_{half}$ plane of Figure \ref{fig:mflat3}.  But we must consider that since the objects with smallest half-light radii tend also to be found closest to the Milky Way, tides may have displaced them systematically farther along the tidal track than larger, more distant systems.  This may have increased the slope of the relation in Figure \ref{fig:mflat3}.  Under this scenario, the dSph population may have formed according to a flatter relation better described by an isothermal sphere ($M\propto r$), but then evolved tidally such that the relation we observe today is the steeper power law $M\propto r^{1.4}$.  However, we re-emphasize that while tides may have increased both the scatter and slope of the empirical correlations, movement of individual systems along tidal tracks would not have introduced correlations where none existed initially.  

\subsection{Sagittarius}

Sagittarius (Sgr) is the only one of the Local Group dSphs that is unambiguously in the throes of tidal disruption \citep{ibata95,mateo96,majewski03}.  The largest of the objects in our sample, the main body of Sgr has a half-light radius that is slightly larger than the scale radius of the best-fitting universal NFW halo.  This is compatible with the fact that Sagittarius continues to lose stars, as its streams of debris wrap more than once around the sky \citep{majewski03}.  Perhaps strangely, Sgr is otherwise inconspicuous in our scaling relations, falling neatly onto the best-fitting power-law profile.  According to P08's tidal tracks, the progenitor of Sgr had larger velocity dispersion than Sgr exhibits today, such that Sgr may originally have been a more extreme outlier than Cetus and Tucana are now. 

\subsection{On the Ultra-Faint Satellites}

Since they extend the range of dSph sizes downward by an order of magnitude, the ultra-faint dSphs provide much of the leverage in discerning the scaling relations underlying Figures \ref{fig:mflat2}-\ref{fig:mflat3}.  Therefore we must note some caveats due to the fact that the least luminous objects are necessarily the least well-studied.  First, their velocity dispersions are the most uncertain, a result of small sample sizes (e.g., just five stars in the case of Leo V \citep{walker09c}) and the fact that their small dispersions are near the resolution limits of the spectrographs used to measure them \citep{simon07}.  In some cases, the measured dispersions (and corresponding masses and densities) represent just upper limits, so the data for objects like Leo V and Bootes II remain consistent with these objects  having tiny dispersions that would render them more similar to star clusters than to bona fide dSph galaxies. 

Second, many (e.g., BooII, Coma, Segue 1, Segue 2, UMaII, Willman 1) of the ultra-faint satellites are found within $\leq$ 50 kpc of the Sun, where they are most vulnerable to Milky Way tides (see also section \ref{subsec:tides}).  While extreme central densities ($\la 5M_{\odot}$pc$^{-3}$) should be sufficient to protect the smallest systems against tidal disruption, many of the ultra-faint satellites appear elongated (e.g., Hercules has axis ratio 3:1 \citep{coleman07}) or otherwise exhibit irregular morphologies (e.g., UMaII has an apparent double core \citep{zucker06b}).  On one hand, one does not expect unbound tidal debris to inflate the velocity dispersions of the smallest dSphs---\citet{penarrubia08c} use simulations to show that unbound debris escapes from a disrupting dSph on a timescale similar to the crossing time, which for the smallest satellites is just $\sim r_{half}/\sigma\sim 10$ Myr.  On the other hand, if distorted morphologies result from strong tidal processes, then it is difficult to understand how these systems could be embedded in intact dark matter halos similar to those inferred for more regular dSphs.  It may turn out, then, that some of the recently discovered satellites are unbound star clusters in the process of dissolving.  In that case, it would be puzzling that these systems manage to fall on or near the empirical relations defined by bona fide dSphs.  

\section{A Universal Mass Profile?}
\label{sec:universal}

We have shown that there exists in the population of known dSphs an empirical correlation between half-light radius and velocity dispersion (Figure \ref{fig:mflat3}) that, under equilibrium models, implies a correlation of the form $M(r_{half})\propto r_{half}^{1.4\pm 0.4}$ (Figure \ref{fig:mflat2}).  We emphasize that this implication is insensitive to the assumption of isotropy inherent in Equations \ref{eq:simplejeansmass}-\ref{eq:mrhalf}---our Jeans/MCMC analysis shows that our estimates of $M(r_{half})$ are robust over a wide range of (constant) anisotropy values.  Taken at face value, these results suggest that the smallest dSphs may simply be embedded more deeply inside dark matter halos that are otherwise similar to those inhabited by the larger (and more luminous) dSphs.  Then let us examine the extent to which dSph dark matter halos might be ``similar.''  In fact let us carry this notion to its limit and consider the hypothesis that all dSphs are embedded within identical dark matter halos.  In that case our robust constraints on $M(r_{half})$ become data points that sample a single, ``universal'' mass profile at discrete radii, thereby allowing us to measure the profile's shape directly and free from the concerns regarding mass/anisotropy degeneracy that plague analyses of individual systems.  We have already shown that the data are broadly consistent with a power-law profile of the form $M(r)\propto r^{1.4}$.  In what follows we also consider cusped and cored profiles, and then we examine the ability of a universal profile to fit the individual velocity dispersion profiles of the brightest dSphs.  Finally, we quantify the amount of empirical scatter with respect to each candidate for a universal dSph profile. 

\subsection{Cusps versus Cores}
\label{subsec:corecusp}

We now investigate whether the correlations in Figures \ref{fig:mflat2} and \ref{fig:mflat3} can be described in terms of universal versions of the halo models considered in Section \ref{sec:jeans}.  For simplicity we restrict our consideration to two specific halos of interest: the cuspy NFW halo with $\alpha=\gamma=1$ and a cored halo with $\alpha=1$, $\gamma=0$.  In order to compare these halo models directly with observable quantities, we equate $M(r_{half})$ from Equation \ref{eq:hernquist2} with the model-independent estimate given by Equation \ref{eq:mrhalf}.  For the NFW halo, this gives the scaling relation
\begin{equation}
  r_{half}\sigma^2=\frac{2\eta r_0V_{max}^2}{5}\frac{\ln[1+r_{half}/r_0]-\frac{r_{half}/r_0}{1+r_{half}/r_0}}{\ln[1+\eta]-\frac{\eta}{1+\eta}},
  \label{eq:nfwsigma}
\end{equation}
where $\eta\sim 2.16$.  For the cored halo, the scaling relation becomes
\begin{eqnarray}
  r_{half}\sigma^2=\frac{2\eta r_0V_{max}^2}{5(\ln[1+\eta]+\frac{2}{1+\eta}-\frac{1}{2(1+\eta)^2}-\frac{3}{2})}\hspace{0.5in}\nonumber\\
  \times \biggl [ \ln[1+r_{half}/r_0]+\frac{2}{1+r_{half}/r_0}-\frac{1}{2(1+r_{half}/r_0)^2}-\frac{3}{2}\biggr ],
  \label{eq:coresigma}
\end{eqnarray}
now with $\eta\sim 4.42$ for $\alpha=1$, $\gamma=0$.  

\begin{figure}
  \epsscale{1.2}
  \plotone{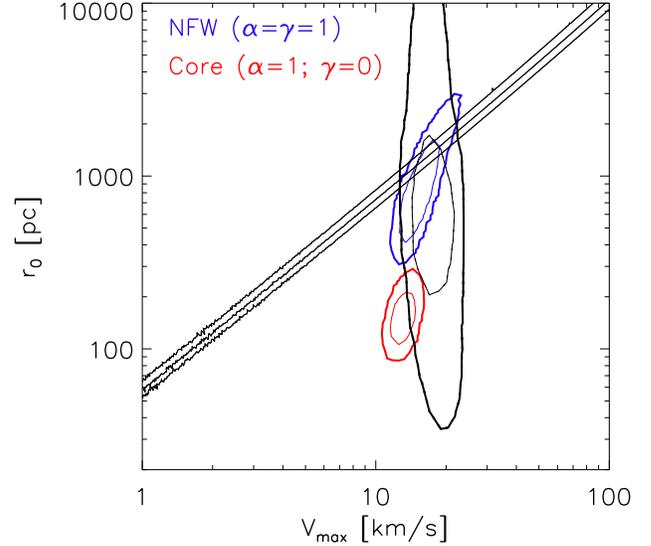}
  \caption{\scriptsize Constraints on halo parameters $V_{max}$ and $r_0$ if we fit a single NFW (blue) or cored (red) halo to the $\sigma_{V_0}$ vs. $r_{half}$ relation for all dSphs.  Contours indicate 68\% and 95\% confidence regions, from a $\chi^2$ fit.  For comparison, solid black contours show model-independent constraints obtained for the Fornax halo alone (Section \ref{subsec:mcmc} and Figure \ref{fig:for_par5}).  Overplotted are the mass-concentration relations for cold dark matter halos at $z=0,1,2$, from cosmological N-body simulations \citep{navarro96,eke01,bullock01}.}
  \label{fig:halo}
\end{figure}

Let us consider whether the correlations in Figures \ref{fig:mflat2} and \ref{fig:mflat3} can be fit with a single halo of either the cusped or cored variety.  Contours in the $V_{max}$-$r_0$ plane in Figure \ref{fig:halo} indicate constraints we obtain by fitting, via Equations \ref{eq:nfwsigma} and \ref{eq:coresigma}, NFW and cored profiles to the $r_{half}$-$\sigma_{V_0}$ data.  Accounting for measurement errors in both dimensions, the best-fitting NFW halo has $V_{max}\sim 15$ km s$^{-1}$, $r_0\sim 795$ pc, while the best-fitting cored halo has $V_{max}\sim 13$ km s$^{-1}$, $r_0\sim 150$ pc.  The best-fitting NFW and cored profiles are plotted over the empirical relations in Figures \ref{fig:mflat2} and \ref{fig:mflat3}.  

For both NFW and cored halos the scale radius is constrained by the slope in the empirical $\sigma$-$r_{half}$ relation (Figure \ref{fig:mflat3}).  Cored halos with scale radii larger than $r_0\ga 200$ pc generate correlations with slopes in the $\sigma$-$r_{half}$ plane that are systematically steeper than that of the empirical relation.   We have tried other forms of the cored halo (e.g., $\alpha=0.5, \alpha=2$) and find that in all cases, the scale radius must be $\la 200$ pc.  Another way of understanding this constraint is that it is set by the fact that the empirical relation in Figure \ref{fig:mflat2} (right panel) does not show signs of turning over as $r\rightarrow 0$.  Thus the universal dSph halo, if it is cored, must have small scale radius.  

Figure \ref{fig:halo} also demonstrates that for a universal NFW halo, the allowed region of parameter space intersects the mass-concentration relation that is derived independently from cosmological N-body simulations \citep{navarro96,eke01,bullock01,penarrubia08a}.  Thus the single NFW halo that best fits all the data is broadly consistent with the $\Lambda$CDM framework.

From both the cusped and cored halo fits there is a clear prediction that deviates from that of the power-law fit: collectively, dSph velocity dispersions increase with half-light radius until reaching a maximum of $\sigma\sim 10$ km s$^{-1}$ for $r_{half}\sim 1$ kpc (Figure \ref{fig:mflat3}).  Larger systems falling on the same relation will have \textit{smaller} velocity dispersions, $\sigma\la 10$ km s$^{-1}$, as the half-light radius becomes similar to the scale radius of the dark matter halo.  We will soon be able to test this prediction as kinematic data become available for more dSph satellites of M31, many of which have $r_{half}\ga 1$ kpc \citep{mcconnachie06}.  The universality of the best-fitting cusped or cored halos would imply that dSphs represent a class of spheroid that is at once distinct from globular clusters ($r_{half}\sim 1-10$pc and $\sigma\sim 5-10$ km s$^{-1}$) and larger dwarf elliptical galaxies ($r_{half}\sim 1$ kpc, $\sigma_{V_0}\sim 50$ km s$^{-1}$)---both types of object have velocity dispersions too large for the dSph relation for cored or cusped halos.  In contrast, a power-law profile would imply that velocity dispersion increases with half-light radius indefinitely (dashed line in Figure \ref{fig:mflat3}), admitting the possibility that dSph mass profiles are similar to those of larger ellipticals and perhaps spiral galaxies (Paulo Salucci, private communication; Stacy McGaugh, private communication).  

Irrespective of whether the halo is cored or cusped, the hypothesis of universality leads to another clear prediction. If the right-hand side of Equation \ref{eq:jeans} is universal, then the left-hand side must be as well. So, there is a universal functional relationship between the dSph stellar density, velocity dispersion and anisotropy.  This is simple to work out in the case of isotropy and constant velocity dispersion. Two dSphs with velocity dispersions $\sigma_1$ and $\sigma_2$ must have luminosity profiles $\nu_1$ and $\nu_2$ related by
\begin{equation}
  \nu_1^{\sigma_1^2} = \nu_2^{\sigma_2^2}
\end{equation}
In other words, the two profiles are related via a power-law transformation.  Even in the more complicated case when the anisotropy $\beta$ is non-zero, the luminosity profiles of different dSphs are still related to each other, though the transformation is now more complicated than a power-law.

\subsection{Velocity Dispersion Profiles}
\label{subsec:vdisp}

A universal dSph dark matter halo must account not only for the bulk kinematic properties of the dSph population, but also for the velocity dispersion profiles, where available, of individual dSphs.  Therefore let us test whether the four candidate profiles considered in this work---a singular isothermal sphere with $M\propto r$, a power-law profile with $M\propto r^{1.4}$, an NFW cusp with scale radius $r_0=795$ pc, and a core ($\alpha=1; \gamma=0$) with $r_0=150$ pc---are consistent with the velocity dispersion profiles in Figure \ref{fig:momentprofiles}.  For each profile we fit to the empirical velocity dispersion profiles using the Jeans equation (Equation \ref{eq:jeansproject}), allowing now for just two free parameters---anisotropy and the normalization, $V_{max}$.  For the power-law profile the circular velocity curve is unbounded as $r\rightarrow \infty$, so we normalize the power-law profile with the value of $M_{300}$.  In order to consider only the most realistic cases, we restrict the anisotropy to values between $-0.6\leq \beta \leq +0.3$.  

For each of the four universal profiles considered, Table \ref{tab:universal} lists the values of $\beta$ and $V_{max}$ that produce the best fits to the empirical velocity dispersion profiles of the bright dSphs.  We find that for each halo profile, the data for all eight dSphs are well fit if we allow the normalization to vary over a factor of $\la 2$.  For the power-law profile, we can fit all the empirical velocity dispersion profiles if we let $M_{300}$ take values between $[0.5-1.5]\times 10^7M_{\odot}$---note that this is the same range for $M_{300}$ found by S08, but unlike in that study, here $M_{300}$ serves as the normalization for mass profiles of the same shape.  The isothermal, cored and NFW profiles require $V_{max}$ in the range $10-20$ km s$^{-1}$.  The best fits for each profile are plotted against the empirical velocity dispersion profiles in Figure \ref{fig:momentprofiles}, and parameters of those fits are listed in Table \ref{tab:universal}. 

\begin{deluxetable*}{lrrrrrrrrrr}
\tabletypesize{\scriptsize}
\tablewidth{0pc}
\tablecaption{Anisotropy and Normalization of Universal Profile, for dSphs in Figure 1}
\tablehead{\\
  &\colhead{$\underline{M\propto r}$}&\colhead{$\underline{M\propto r^{1.4}}$}&\colhead{\underline{NFW ($r_0=795$pc})}&\colhead{\underline{Core ($r_0=150$pc})}\\
  \colhead{Galaxy}&\colhead{$\beta;\hspace{0.05in} V_{max}$}&\colhead{$\beta;\hspace{0.05in} M_{300}$}&\colhead{$\beta;\hspace{0.05in} V_{max}$}&\colhead{$\beta;\hspace{0.05in} V_{max}$}\\
  \colhead{}&\colhead{\hspace{0.3in}$[M_{\odot}]$}&\colhead{\hspace{0.3in}$[M_{\odot}]$}&\colhead{\hspace{0.3in}[km s$^{-1}$]}&\colhead{\hspace{0.3in}[km s$^{-1}$]}\\
}
\startdata
Carina&$-0.6^{+0.5};\hspace{0.05in} 10_{-1}^{+1}$&$-0.6^{+0.9};\hspace{0.05in} 0.7_{-0.2}^{+0.2}\times 10^7$&$-0.4_{-0.2}^{+0.6};\hspace{0.05in} 13_{-1}^{+1}$&$-0.5_{-0.1}^{+0.7};\hspace{0.05in} 11_{-1}^{+2}$\\
\\
Draco&$-0.5_{-0.1}^{+0.5};\hspace{0.05in} 13_{-1}^{+4}$&$-0.6^{+0.9};\hspace{0.05in} 1.6_{-0.4}^{+0.6}\times 10^7$&$-0.6^{+0.8};\hspace{0.05in} 20_{-2}^{+3}$&$-0.6^{+0.7};\hspace{0.05in} 17_{-2}^{+3}$\\
\\
Fornax&$-0.6^{+0.2};\hspace{0.05in} 16_{-1}^{+1}$&$-0.2_{-0.4}^{+0.3};\hspace{0.05in} 1.2_{-0.1}^{+0.1}\times 10^7$&$-0.3_{-0.3}^{+0.3};\hspace{0.05in} 18_{-1}^{+1}$&$-0.5_{-0.1}^{+0.1};\hspace{0.05in} 16_{-0}^{+2}$\\
\\
Leo I&$-0.6^{+0.7};\hspace{0.05in} 16_{-4}^{+2}$&$-0.6^{+0.9};\hspace{0.05in} 1.6_{-0.3}^{+0.4}\times 10^7$&$-0.2_{-0.3}^{+0.6};\hspace{0.05in} 20_{-2}^{+2}$&$-0.6^{+0.9};\hspace{0.05in} 18_{-2}^{+2}$\\
\\
Leo II&$-0.4_{-0.1}^{+0.8};\hspace{0.05in} 12_{-2}^{+2}$&$+0.3_{-0.9};\hspace{0.05in} 1.2_{-0.5}^{+0.5}\times 10^7$&$+0.3_{-0.9};\hspace{0.05in} 18_{-3}^{+3}$&$+0.3_{-0.9};\hspace{0.05in} 15_{-3}^{+2}$\\
\\
Sculptor&$-0.6^{+0.2};\hspace{0.05in} 14_{-1}^{+1}$&$-0.5_{-0.1}^{+0.3};\hspace{0.05in} 1.1_{-0.3}^{+0.4}\times 10^7$&$-0.5_{-0.1}^{+0.5};\hspace{0.05in} 19_{-1}^{+1}$&$-0.6^{+0.4};\hspace{0.05in} 16_{-1}^{+1}$\\
\\
Sextans&$-0.5_{-0.1}^{+0.2};\hspace{0.05in} 14_{-1}^{+1}$&$-0.5_{-0.1}^{+0.7};\hspace{0.05in} 0.5_{-0.2}^{+0.2}\times 10^7$&$-0.4_{-0.2}^{+0.6};\hspace{0.05in} 11_{-2}^{+2}$&$-0.5_{-0.1}^{+0.6};\hspace{0.05in} 10_{-2}^{+1}$\\
\\
UMi&$-0.6^{+0.5};\hspace{0.05in} 14_{-2}^{+2}$&$-0.6^{+0.7};\hspace{0.05in} 1.3_{-0.3}^{+0.4}\times 10^7$&$-0.6^{+0.8};\hspace{0.05in} 19_{-2}^{+2}$&$-0.6^{+0.7};\hspace{0.05in} 15_{-2}^{+2}$\\
\enddata
\label{tab:universal}
\end{deluxetable*}

\subsection{Scatter}

\begin{figure}
  \epsscale{1.2}
  \plotone{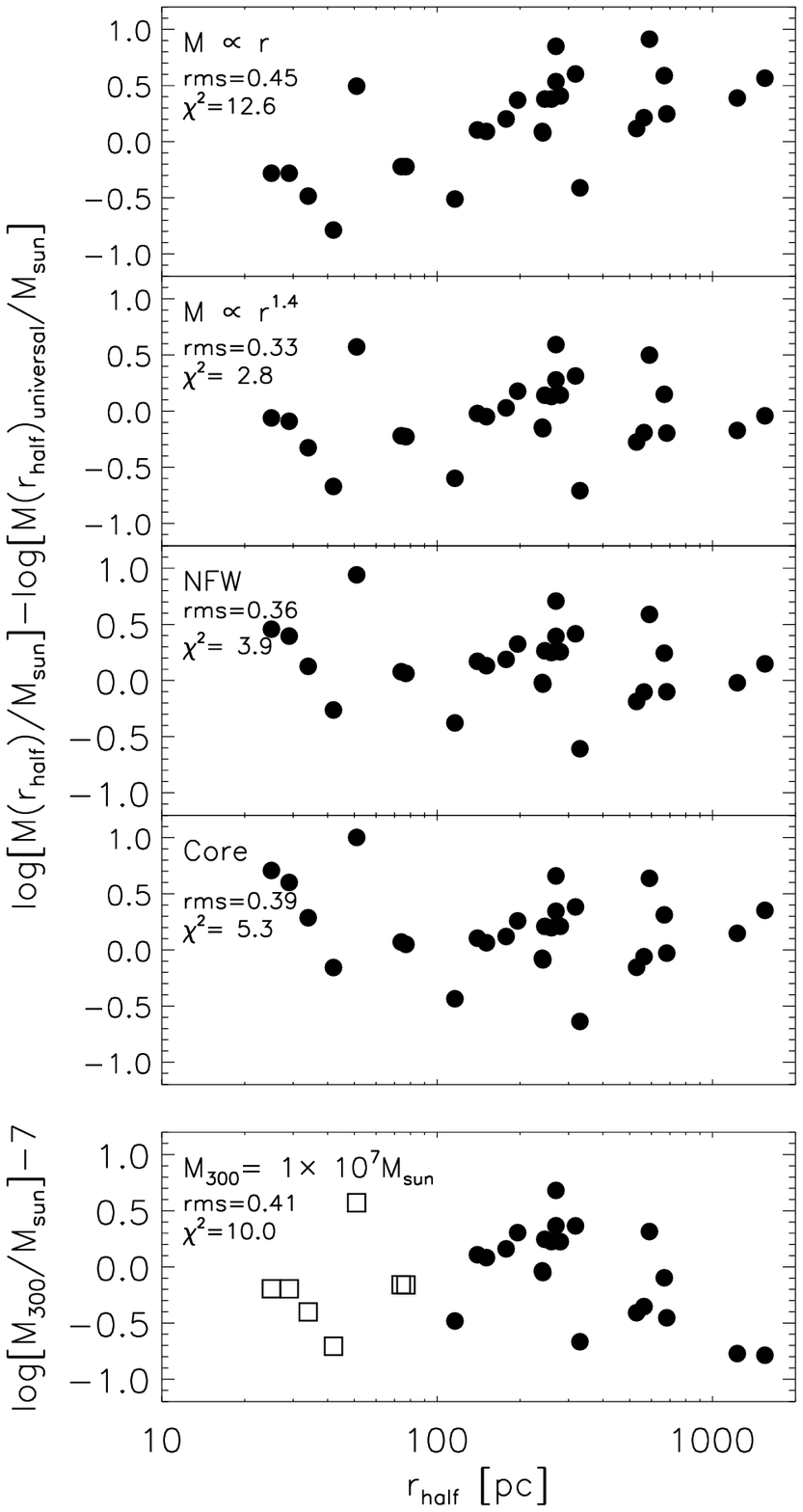}
  \caption{\scriptsize \textit{Top four panels:} Residuals with respect to universal mass profiles considered in this work.  \textit{Bottom:} Residuals with respect to the common mass scale $M_{300}=10^7M_{\odot}$.  Empirical masses are estimated from Equations \ref{eq:simplejeansmass}-\ref{eq:mrhalf}.  In the bottom panel, open squares represent ultra-faint satellites for which $r_{half}< 100$ pc, and for which there is no available evidence that a dark matter halo actually extends to a radius of 300 pc.}
  \label{fig:mflat_residuals}
\end{figure}

We now evaluate the feasibility of a universal dSph mass profile in terms of the amount of scatter with respect to the best candidates for that profile.  Figure \ref{fig:mflat_residuals} plots residuals $\Delta (\log M)\equiv \log [M(r_{half})/M_{\odot}]-\log [M(r_{half})_{universal}/M_{\odot}]$ (where the first term represents the data and the second term represents the universal profile) with respect to the best-fitting isothermal, power-law, NFW and cored profiles considered in this work.  For comparison the bottom panel plots the residuals in the common-$M_{300}$ relation, $\log [M_{300}/M_{\odot}]-7$, where $M_{300}$ is estimated from Equation \ref{eq:simplejeansmass}.  Text indicates for each profile the statistic $\chi^2=N^{-1}\sum_{i=1}^N(\Delta (\log M))^2/\delta^2$, where $\delta\equiv\sigma_{M(r_{half})}[M(r_{half})\ln(10)]^{-1}$ represents the measurement error in log-space.  The power-law ($M(r)\propto r^{1.4}$) provides the best fit ($\chi^2=2.8$), followed by the NFW profile ($\chi^2=3.9$), the cored profile ($\chi^2=5.3$), and the isothermal profile ($\chi^2=12.6$).  While these values of $\chi^2$ indicate scatter larger than what is expected from the quoted errors if the profile is truly universal, we note that that the scatter of $M_{300}$ values about a common mass of $10^7M_{\odot}$ corresponds to $\chi^2=10.0$.  In addition to $\chi^2$, Figure \ref{fig:mflat_residuals} also indicates the root-mean-square of the residuals.  There we also find that the scatter about a universal mass profile ($rms=0.33$ for the power-law profile) is similar to the scatter about a common value of $M_{300}$ ($rms=0.41$).

Thus in terms of empirical scatter, the hypothesis that dSphs follow a universal mass profile receives as much support from the data as the claim that all dSphs have a common $M_{300}$.  Note that the former is the stronger claim, since it implies the latter (not vice-versa), and both statements involve an extrapolation of sorts.  S08's argument for a common $M_{300}$ rests on extrapolation, since for the smallest objects ($r_{half}\sim 30$ pc) it requires evaluation of mass profiles at radii that are larger by an order of magnitude than the region containing kinematic data.  In contrast the claim for a universal mass profile emerges from the pattern of masses robustly measured at the half-light radius of each dSph.  Yet it is important to acknowledge that we know the enclosed mass only at this one particular radius in each dSph, so we cannot definitively rule out the situation in which any given dSph has a mass profile that diverges wildly from the hypothesized universal profile at radii far from its own half-light radius.  However, if this circumstance were true for more than just a few outliers, then it would require a remarkable coincidence for the masses at dSph half-light radii to generate the appearance of a universal profile over two orders of magnitude in $r_{half}$.  We conclude that if one is impressed by the commonality of $M_{300}$ among dSphs, one is obliged to be impressed by the apparent commonality of dSph masses at all radii.  

\section{Discussion and Summary}

In the first part of our analysis (Section \ref{sec:jeans}), we have applied the Jeans equation to the velocity dispersion profiles of eight bright dSphs.  The quality of the data set now available for Fornax allows us to place the first model-independent constraint on the maximum circular velocity of a dwarf spheroidal.  While for all other galaxies we obtain only lower limits of $V_{max}\ga 10$ km s$^{-1}$, for Fornax our Jeans/MCMC anlysis yields $V_{max}=18_{-3}^{+5}$ km s$^{-1}$.  This constraint enables straightforward comparison to subhalos produced in cosmological N-body simulations, for which $V_{max}$ is readily quantifiable, and motivates the pursuit of larger velocity samples in the remaining dSphs. 

A key result from our Jeans/MCMC analysis is that for all well-observed dSph galaxies with measured velocity dispersion profiles, most allowed mass profiles intersect near the half-light radius, regardless of the assumed anisotropy.  This result is consistent with previous findings by \citet{strigari08} and \citet{penarrubia08a}, and has two important consequences.  First, while the Jeans analysis does not place meaningful constraints on most halo parameters for most galaxies, for all eight ``classical'' dSphs we obtain relatively tight constraints on the enclosed mass at the half-light radius.  Second, the insensitivity of $M(r_{half})$ to anisotropy means that one need not employ the full Jeans/MCMC analysis to obtain reliable estimates of $M(r_{half})$.  In Section \ref{subsec:simple} we derive Equations \ref{eq:simplejeansmass} and \ref{eq:mrhalf} from the simplifying assumptions that dSph stellar components have isotropic velocity distributions with constant dispersion.  These mass estimators have the advantage that they require as input only measurements of the velocity dispersion and half-light radius, and so can be applied to the faint dSphs for which velocity dispersion profiles are unavailable.  Using Equation \ref{eq:simplejeansmass} to estimate $M_{300}$, we immediately reproduce the flat $M_{300}$-luminosity relation (Figure \ref{fig:m300}) originally obtained by the Jeans/MCMC analysis of S08.  Using Equation \ref{eq:mrhalf}, we obtain estimates of $M(r_{half})$ that stand in excellent agreement with the results of our own Jeans/MCMC analysis (Figure \ref{fig:all_mrhalfbeta} and Table \ref{tab:mcmc}) for the brightest dSphs.   

In the second part of this work, we have applied the simple estimator of $M(r_{half})$ given by Equation \ref{eq:mrhalf} to the entire dSph population in lieu of the full Jeans/MCMC analysis.  We find that the 28 dSphs with published kinematic data exhibit a correlation of the form $M(r_{half})\propto r_{half}^{1.4\pm 0.4}$ (Figure \ref{fig:mflat2}, left panel).  This slope is steeper than that of the singular isothermal sphere because, as the inclusion of the ultra-faint dSphs makes clear, velocity dispersion increases with half-light radius (Figure \ref{fig:mflat3}).  Constraints on $M(r_{half})$ imply constraints on the mean density interior to the half-light radius, and the smallest dSphs now allow us to measure galactic densities at the scale of tens of parsecs.  Toward these small radii, the mean densities we measure continue to rise according to the power law that best fits the empirical relation, $\langle \rho\rangle\propto r^{-1.6\pm 0.4}$ (Figure \ref{fig:mflat2}, right-hand panel).  For the smallest systems, this behavior may imply extremely large densities of order $\langle \rho \rangle \sim 5M_{\odot}$pc$^{-3}\sim 200$Gev cm$^{-3}$.  However, we emphasize that for some of the smallest systems, the available data do not place firm lower limits on velocity dispersion, so these densities should be viewed as upper limits.  

We have argued in Section \ref{subsec:tides} that tidal forces have not introduced the scaling relations underlying Figures \ref{fig:mflat2}-\ref{fig:mflat3}.  According to the tidal tracks provided by the N-body simulations of \citet{penarrubia08b}, the long-term effect of tidal stripping is to reduce a dSph's velocity dispersion faster than its size.  This moves a dSph off the empirical $\sigma$-$r_{half}$ relation rather than along it.  Thus tides act primarily to increase the scatter in the scaling relations we observe today.  We conclude that the correlation between velocity dispersion and $r_{half}$, and the resulting relation between $M(r_{half})$ and $r_{half}$, are inherent to the dSph formation mechanism.  However, tides may have altered the slope of the empirical correlations we observe today.  Since the dSphs closest to the Milky Way tend also to be the smallest and least luminous, the stronger tides they have encountered may have shifted their velocity dispersions systematically downward, thereby increasing the slope in the $\sigma$-$r_{half}$ relation.  

Finally, in Section \ref{sec:universal} we have considered, in light of the correlation between $M(r_{half})$ and $r_{half}$, the possibility that all dSphs follow a universal mass profile.  We have shown that in addition to the simple power law $M(r)\propto r^{1.4}$, the empirical relationship between mass and half-light radius (or equivalently, between velocity dispersion and half-light radius) can be fit by an NFW ($V_{max}=15$ km s$^{-1}$, $r_0=795$ pc) or cored ($V_{max}=13$ km s$^{-1}$, $r_0=150$ pc) halo profile.  The allowed parameters of the universal NFW halo (Figure \ref{fig:halo}) overlap the mass-concentration relation derived from cosmological N-body simulations, supporting the internal consistency of the $\Lambda$CDM framework.  While we obtain the best fits to the velocity dispersion profiles of the brightest dSphs if we allow the normalizations of the profiles to vary over a factor of $\la 2$ (Section \ref{subsec:vdisp}), we find that the empirical scatter of the $M(r_{half})-r_{half}$ data about a universal profile is similar to the scatter about a common value of $M_{300}$.  Thus the stronger claim that dSphs follow a universal mass profile receives at least as much support from the data as the independent claim that all dSphs have a common $M_{300}$.  
 
It has long been suggested that dSphs exhibit a common mass scale.  Fifteen years ago, this conclusion followed simply from the virial theorem and the fact that the bright dSphs known at the time ($L_V\ga 10^5L_{V,\odot}$) all have similar sizes and velocity dispersions \citep{mateo93}.  The recently-discovered ultra-faint dSphs, which are systematically smaller and kinematically colder than the brighter dSphs, undermine a common dSph mass scale as derived from the virial theorem.  \citet{simon07} show that the ultra-faint dSphs deviate from the luminosity-$M/L$ relation followed by the brighter dSphs \citep{mateo98}, and the dynamical masses of the smallest dSphs are just $\sim 10^5M_{\odot}$ (e.g., \citealt{geha08,walker09c}).  S08 are able to recover a common dSph mass scale of $M_{300}\sim 10^7M_{\odot}$ that includes all ultra-faint as well as bright dSphs, but doing so requires them to evaluate all mass profiles at a fixed radius that happens to lie well beyond the stellar component in the smallest dSphs.  In contrast, we have presented the case for a common dSph mass profile that emerges from masses that relate directly to empirical dynamical properties---tracer density and velocity dispersion---in the regions where they are directly constrained by data, requiring no extrapolation to radii beyond the optical extent of the galaxy.

We thank Vasily Belokurov and Jaroslaw Klimentowski for providing comments and helpful discussions regarding this work.  We thank Lindsay King, Damien Quinn and Antony Lewis for assistance regarding the application of MCMC techniques.  We also thank the anonymous referee for providing criticism that helped improve this work.  MGW and JP acknowledge support from the STFC-funded Galaxy Formation and Evolution programme at the Institute of Astronomy, University of Cambridge.  MM acknowledges support from NSF grants AST-0507453, and 0808043.  EO acknowledges support from NSF Grants AST-0505711, and 0807498.

\bibliography{ref}
\end{document}